\documentclass[twocolumn,showpacs,amsmath,amssymb,nofootinbib]{revtex4}
\usepackage{graphicx} 
\usepackage{dcolumn}

\begin{document}
\title{Coexistence of '$\alpha$+ $^{208}$Pb' 
cluster structures and single-particle excitations 
in $^{212}_{~84}$Po$^{}_{128}$}

\author{A. Astier$^1$}
\author{P. Petkov$^{1,2}$}
\author{M.-G. Porquet$^1$}
\author{D.S. Delion$^{3,4}$}
\author{P. Schuck$^5$}

\affiliation{$^1$CSNSM, IN2P3-CNRS and Universit\'e Paris-Sud 
91405 Orsay, France\\
$^2$INRNE, BAS, 1784 Sofia, Bulgaria\\
$^3$Horia Hulubei National Institute of Physics and Nuclear 
Engineering  407 Atomistilor, 077125 Bucharest, Romania\\
$^4$ Academy of Romanian Scientists, 54 Splaiul Independentei  
050094 Bucharest, Romania\\
$^5$ IPN, IN2P3-CNRS and Universit\'e Paris-Sud  91406 Orsay, France\\
}

\begin{abstract}
Excited states in $^{212}$Po have been populated by $\alpha$
transfer using the $^{208}$Pb($^{18}$O, $^{14}$C) reaction  at 85~MeV beam 
energy and studied with the EUROBALL~IV $\gamma$ multidetector array.
The level scheme has been extended up to $\sim 3.2$~MeV excitation energy 
from the triple $\gamma$ coincidence data. Spin and parity values
of most of the observed states have been assigned from   
the $\gamma$ angular distributions and $\gamma -\gamma$ angular correlations.
Several $\gamma$ lines with E$_\gamma <$ 1~MeV have been found to be 
shifted by the Doppler effect, allowing for the 
measurements of the associated lifetimes by the DSAM method. The values, found 
in the range [0.1-0.6]~ps, lead to very enhanced E1 transitions. All the
emitting states, which have non-natural parity values, are discussed in 
terms of $\alpha$-$^{208}$Pb structure. They are in the same 
excitation-energy range as the states issued from shell-model configurations. 
\end{abstract}


\maketitle
\section{Introduction}\label{introduction}

The $^{212}$Po nucleus is the last member of the $^{232}$Th radioactive chain. 
As early as 1916, Rutherford and Wood discovered ~\cite{ru16} the
emission of $\alpha$ particles from the end of this chain, 
the "active deposit" of
thorium (formerly named as ThB). Many years later, this was interpreted
as the $\alpha$ decays of several states of $^{212}$Po, particularly
its ground state which is a pure $\alpha$ emitter, 
with a very short half-life, 0.299(2)~$\mu$s. Afterwards new $\alpha$ activities were
produced by heavy/light-ion bombardments of bismuth and lead, and
their decay spectra were measured~\cite{pe62}. These studies have shown that an 
$\alpha$-line at 11.6~MeV arises directly from an isomeric state of 
$^{212}$Po with a half-life of 45~s, i.e. $1.5 \times 10^8$ longer than the ground-state
one. The very large hindrance factor for this $\alpha$ emission was explained by
spin and parity assignments of 18$^+$ for the $^{212}$Po isomeric state, which has
been located at 2.9~MeV excitation energy. 

Besides these two extreme states, other levels were identified from the 
$\beta$-decay of the two long-lived states
of  $^{212}$Bi or {\it via} light-ion induced transfer reactions, such as 
($^{18}$O,$^{14}$C) or ($^9$Be,$\alpha$ n) on a $^{208}$Pb
target~\cite{bo81,po87}, and all the yrast levels were discussed in terms 
of shell model configurations.
The $^{212}$Po nucleus has been also studied in neutron-evaporation channels 
from total-fusion reactions, using either a radioactive target~\cite{su85}, 
$^{210}$Pb($\alpha$, 2n), or a radioactive beam~\cite{po03,ga05}, 
$^{208}$Pb($^8$He,4n). However, the limited number of the target 
nuclei or the projectiles did not sizeably improve the 
spectroscopic data obtained in the $^{208}$Pb($^9$Be,$\alpha$ n)
reaction~\cite{po87}.

There were numerous attempts to give a microscopic description of the $\alpha$ 
decay of the ground state of $^{212}$Po. The use of shell-model configurations 
fails to reproduce the large value of its rate by an order of magnitude, meaning 
that the $\alpha$-particle formation is badly predicted within this approach. 
When complemented by cluster-model-type basis states~\cite{va92}, the decay 
width is predicted in good agreement with the experimental value, 
the amount of $\alpha$ clustering in the ground state of $^{212}$Po being $\sim$0.3.
This is not far from the high degrees of clustering well known in some light 
nuclei where the two neutrons and the two protons occupy the same orbit, which is not
fulfilled in the present case.
 
The existence of the $\alpha$ cluster would also act upon the properties of the  
excited states of $^{212}$Po. Whereas shell-model approaches
work well to predict the excitation energy of the yrast states~\cite{po87}, they
cannot reproduce their large experimental B(E2) transition strengths, 
which are well accounted for from the $\alpha$ clustering 
(see for instance refs.~\cite{bu94,ho94,oh95,bu96}).
All these results point to the fact that the positive-parity yrast states do 
contain both components, shell-model configurations and $\alpha$ clustering, 
which are not easy to disentangle. 

In this paper, we present a new class of 
excited states of $^{212}$Po, which have non-natural parity and are 
connected by enhanced E1 transitions to the yrast states. Low-lying
shell-model excitations do not lead to such non-natural parity states and  
these states are the first 
experimental evidence of "pure" $\alpha$-cluster states in heavy nuclei.
The $^{212}$Po nuclei have been obtained using the ($^{18}$O,$^{14}$C) 
transfer reaction on a thick $^{208}$Pb target and
$\gamma ^n$ measurements have been performed. The experimental 
conditions and the various analysis techniques 
are presented in section \ref{experiment}. Section \ref{results} is devoted 
to the experimental results which comprise 
{\it i)} the analysis of the reaction mechanism leading to the $^{212}$Po channel,
{\it ii)} the new level scheme of $^{212}$Po,
{\it iii)} the lifetime measurements using the Doppler Shift Attenuation
Method (DSAM) , 
and {\it iv)} the $\alpha$ branching ratios of the yrast states. 
In the last part (section~\ref{discussion}) we review several properties of 
the first excited states of $^{212}$Po in order to underline the contribution 
of the $\alpha$-clustering. Then we discuss the oscillatory 
motion of the $\alpha$-core distance which may give rise to 
non-natural parity states.

A letter highlighting parts of the present study has been recently
published~\cite{as10}.

\section{Experimental details}\label{experiment}
\subsection{Reaction, $\gamma$-ray detection and analysis}\label{analysis1}
Excited states in $^{212}$Po were populated by $\alpha$ transfer using the 
$^{18}$O$+ ^{208}$Pb reaction, the $^{18}$O beam with an energy of 85~MeV 
being provided by the Vivitron tandem of IReS (Strasbourg). 
A 100~mg/cm$^2$ self-supporting target of $^{208}$Pb was employed, which was 
thick enough to stop the recoiling $^{212}$Po nuclei.
It is worth noting that due to this large thickness, the $^{18}$O beam was
stopped in the Pb target too, and therefore the incident energy covered 
a large range of values, above and below the Coulomb barrier. 
The kinematics of the $\alpha$ transfer 
leading to high-spin states in $^{212}$Po in this experiment 
is detailed below (see sect.~\ref{kinematics}).
The de-exciting $\gamma$-rays  were recorded with 
the EUROBALL IV array consisting of 71 Compton-suppressed Ge 
detectors~\cite{Euroball} 
(15 cluster germanium detectors placed in the backward hemisphere with respect to the beam, 
26 clover germanium detectors located around 90$^\circ$, 30 tapered single-crystal 
germanium detectors located at forward angles).
Each cluster detector is composed of seven closely packed large-volume Ge 
crystals \cite{cluster}
and each clover detector consists of four smaller Ge 
crystals~\cite{clover}.
The 239 Ge crystals of the Euroball array could be grouped into 13 rings, with 
the following angles with respect to the beam
axis, 15.5$^\circ$ (5 crystals), 34.6$^\circ$ (10), 
52.3$^\circ$ (15), 72.2$^\circ$ (26), 80.9$^\circ$ (26),
99.1$^\circ$ (26), 107.5$^\circ$ (26), 122.6$^\circ$ (10), 
130.5$^\circ$ (30), 138.7$^\circ$ (25), 148.1$^\circ$ (15), 
155.9$^\circ$ (15), and 163.5$^\circ$ (10), i.e. 
3 rings forward, 4 rings close to 90$^\circ$ and 6 rings backward 
with respect to the beam axis. 

Events were recorded on tape when at least 3 unsuppressed Ge
detectors fired in prompt coincidence. In this way,
a set of $\sim 4\times 10^9$ three- and higher-fold events were
available for subsequent analysis, but only a small part of these data 
corresponds to $^{212}$Po events. Indeed the main objective of the experiment 
was actually the study of the fusion-fission channel which leads to the 
production of the high-spin states of $\sim 150$ fragments, mainly 
located on the neutron-rich side of the valley of stability~\cite{po04}.
The $^{212}$Po study became itself a goal when it turned out that its main 
$\gamma$ lines were strong enough in our data set to be precisely analyzed. 
We have estimated that the cross section of the exit channel leading 
to $^{212}$Po is $\sim$ 10-20~mb, i.e. $\sim$ 10-20\% of the 
cross section of the fusion-fission channel.

Various procedures have 
then been used for the offline analysis in order to fully 
characterize the excited levels of $^{212}$Po (excitation energy, 
spin and parity values, decay modes, and multipole matrix elements).
Both multi-gated spectra and three-dimensional
'cubes' have been built and analyzed with the 
Radware package \cite{ra95}, starting from the known transitions deexciting
the yrast states~\cite{po87}. 

\subsection{ $\gamma$ angular distributions and $\gamma$-$\gamma$ angular correlations}

\begin{figure}[!h]
\begin{center}
\resizebox{0.4\textwidth}{!}{\includegraphics*{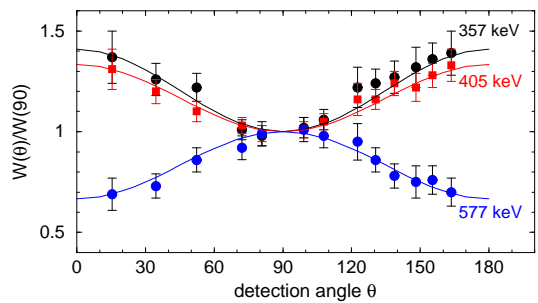}}
\caption{(color online) Angular distribution of some yrast transitions in the $^{212}$Po nucleus
produced in the $^{208}$Pb($^{18}$O,$^{14}$C) reaction. Solid lines are the fits
using the standard Legendre polynomials.
}
\label{distrib_yrast}      
\end{center}
\end{figure}

We have checked whether the reaction mechanism involved here leads to 
spin alignments of $^{212}$Po with respect to the beam axis.
Thus we have chosen several yrast transitions which are 
known to be quadrupole ones with $\Delta I = 2$ (such as the 727, 
405, and 357~keV transitions) or dipole ones with $\Delta I = 1$ (such 
as the 577~keV one). Their angular distributions have been found to be 
symmetric around 90$^\circ$ (see examples in fig.\ref{distrib_yrast}), the intensity of the quadrupole transitions 
being maximum along the beam axis, whereas 
the one of the dipole transition is maximum at 90$^\circ$, demonstrating that  
the spins are aligned in a plane perpendicular to the beam axis.
When fitted
using standard Legendre polynomials $P_{2,4}(cos^2\theta)$, the obtained 
values of the angular coefficient $A_2$/$A_0$ indicate that the alignment
is partial, with an attenuation coefficient $\alpha _2 \sim$ 0.7~\cite{ya67}.

This allowed us to analyze each gamma-ray angular 
distribution to determine its multipole order.
In order to characterize the transitions having too 
weak intensity to be analyzed in that way, their anisotropies have been 
determined using the intensities measured at
two angles relative to the beam axis, 
R$_{ADO}$=I$_{\gamma}$(39.3$^\circ$)/ I$_{\gamma}$(76.6$^\circ$), these two angles
being the average angle of the tapered and cluster detectors in the one hand, 
and the average angle of the clover detectors in the other hand, when taking 
into account the symmetry of the distribution around 90$^\circ$. 
Assuming the same attenuation
coefficient as mentioned above ($\alpha _2 \sim$ 0.7) 
theoretical ADO ratios of 0.85 and 1.30 are expected for 
pure stretched dipole and quadrupole 
transitions, respectively. 

The analysis of $\gamma$-$\gamma$ angular correlations of the most intense 
transitions has been also performed, in order to get rid of the spin alignment 
process. For that purpose, 
the coincidence rates of two successive $\gamma$ transitions are 
analyzed as a function of $\Theta$, the
average relative angle between the two fired detectors.
The Euroball~IV spectrometer had $C^{2}_{239}$~=~28441 
combinations of 2 crystals, out of which only $\sim$ 2000  
involved different values of relative angle within 2$^\circ$. 
Therefore, in order 
to keep reasonable numbers of counts, all the angles 
(taking into account the symmetry around 90$^\circ$) have been 
gathered around three average relative angles : 22$^\circ$, 46$^\circ$, 
and 75$^\circ$ . 

The coincidence rate is increasing between 0$^\circ$ and 
90$^\circ$ for the dipole-quadrupole cascades, whereas it decreases for 
the quadrupole-quadrupole 
or dipole-dipole ones. More precisely, the angular correlation functions 
at the three angles 
of interest were calculated for several combinations of spin sequences 
(see table \ref{correl_th}),
corresponding to typical multipole orders (quadrupole, dipole) and 
typical spin changes ($\Delta I = 2, 1, 0$). 
In order to check the method, angular correlations of transitions 
belonging to the yrast 
cascades of the fission fragments produced in this experiment and 
having well-known multipole orders 
were analyzed and the expected values were found in all cases~\cite{as06}.  
\begin{table}[!h]
\caption[h]{\label{correl_th} Values of the angular correlation functions expected 
for several combinations of spin values, normalized to the ones calculated at 75$^\circ$:
For each $I_1 \to I_2 \to I_3$ sequence given in the first column, the 
spin changes due to the two consecutive transitions
are given in the second column.}
\begin{center}
\begin{tabular}{|cc|ccc|}
\hline
$I_1 \to I_2 \to I_3$ 	&$\Delta I$ & R(22$^\circ$)  & R(46$^\circ$)  & R(75$^\circ$)   \\
 sequence		&values  &   &   &    \\
\hline
6-6-4&0-2  &    1.23  &  1.12  &  1.00 \\
6-4-2&2-2  &    1.13  &  1.06  &  1.00 \\
5-4-3&1-1  &    1.06  &  1.03  &  1.00 \\
5-4-2&1-2  &    0.92  &  0.96  &  1.00 \\
7-6-6&1-0  &    0.91  &  0.95  &  1.00 \\
\hline
\end{tabular}
\end{center}
\end{table}
\subsection{Linear polarization}

Linear polarization can be measured by the use of segmented
detectors acting as a Compton polarimeter, such as the clover detectors of the
Euroball~IV array~\cite{jo95}. 
We have performed such an analysis and have confirmed that 
the three most intense transitions of $^{212}$Po have an electric nature. 
Unfortunately the analysis of the transitions with weaker 
intensity (even those belonging to the yrast band) could not be done. 
While single-gated spectra are useless because of the large background,  
the statistics of the peaks in the double-gated clover spectra 
are too low to precisely measure the effect we are looking for (the
counts in the 'Vertical' and the 'Horizontal' spectra have to differ 
by less than 10\%). 
We have to stress that in this experiment, the major part of the $\gamma$-rays 
are emitted by the fragments of the fusion-fission exit channel, the summed 
multiplicity of their $\gamma$-ray cascades being around 20. On the other hand, 
the $^{212}$Po exit channel has a low $\gamma$-multiplicity. Such a situation 
is not as favourable as that of a rotational structure typically studied in a
fusion-evaporation reaction, even weakly populated, since the high folds of 
their events strongly enhance the statistics of the multi-gated spectra. 

\subsection{Lifetime measurements}
The stopping time of $^{212}$Po in the Pb target is about one 
picosecond, thus it would have been expected that all the 
transitions lying 
in the low-energy part of the level scheme are emitted at rest.
Nevertheless we have found several $\gamma$-rays with E$_\gamma$ $<$
1~MeV which exhibit shifts and broadenings in energy due to the 
Doppler effect, meaning that they are 
emitted in flight and thus the corresponding excited states do have 
lifetimes shorter than 1~ps. 

The lifetime determination using the Doppler-shift attenuation method (DSAM)  
is based on the time-correlation between the slowing-down of the recoiling ion
and the decay of the nuclear level of interest (cf. e.g. 
ref.~\cite{al78}). At a fixed direction of observation, e.g. at an 
angle $\theta$ with respect to the beam axis, the spectrum (line-shape) 
of the registered $\gamma$-rays is given, because of the relation 
$E^{Sh}_{\gamma}$ = $E_{\gamma_0}(1+v_{\theta}/c)$,
by the following formula:
\begin{equation}\label{general formula}
   S_{ij}(E_\gamma) = b_{ij}\int\limits_{-\infty}^{\infty}dE^{Sh}_{\gamma}
       \Phi(E_\gamma,E^{Sh}_{\gamma}) \int\limits_{0}^{\infty} dt
              P_\theta(t,v_{\theta}) \lambda_i n_i(t)
\end{equation}
Here, $v_{\theta}$ is the velocity projection on the observation axis,
the function $n_i(t)$ represents the time-dependent population of the 
level $i$, $\lambda_i$ is its decay
constant and $b_{ij}$ is the branching  of the de-exciting
transition $i$ $\rightarrow$ $j$. The recorded $\gamma$-ray spectrum is 
obtained by a convolution (folding) with the response function $\Phi$ 
of the detector. The quantity $P_\theta(t,v_{\theta})$ is
the stopping matrix which represents the normalized distributions of
the velocity projection $v_{\theta}$ at different times $t$. 
Depending on the experimental situation, additional corrections for geometry, 
efficiency, angular correlation and kinematics effects 
(cf. ref.~\cite{al78}) have also to be taken into account in
Eq.~\ref{general formula}. To reproduce  the line-shape and derive 
the lifetime, it is necessary to correctly determine 
$P_\theta(t,v_{\theta})$, which describes the stopping process, and to
solve Eq.~\ref{general formula} with respect to the decay function
$\lambda_i n_i(t)$  of the level of interest $i$. Thereby,
the influence of the feeding  i.e. of the cascade history has also to be 
considered.

For the description of the slowing-down process via Monte-Carlo methods
(i.e. for the calculation  of the  matrix $P_\theta(t,v_{\theta})$)
we used a modified version of the computer code 
DESASTOP~\cite{wi83a,wi83b} by G. Winter (cf. also ref~.\cite{pe98}).

\section{Experimental results}\label{results}

\subsection{Kinematics of the reaction}\label{kinematics}

\begin{figure}[!h]
\begin{center}
\resizebox{0.35\textwidth}{!}{\includegraphics*{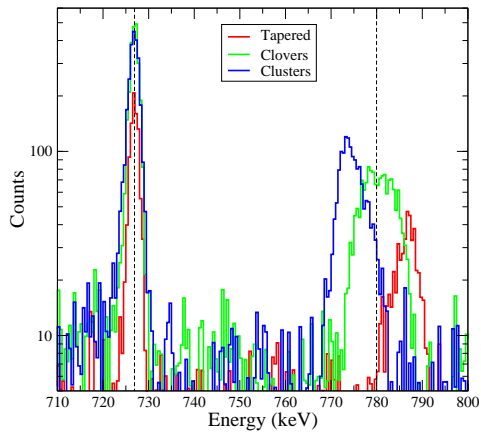}}
\caption{(color online) Parts of typical triple-gated spectra 
as a function of the detection angles  
-tapered Ge (forward angles), clover Ge (around $90^\circ$) 
and cluster Ge (backward angles)- 
emphasizing the Doppler shift and broadening of the 780 keV transition, 
whereas the energy and the width of the 727~keV line do not depend on 
the detection angle.
The two dashed lines indicate the 727.1~keV and 780.4~keV energies.
 }
\label{gate_587keV_TQC}       
\end{center}
\end{figure}

The Doppler effect, if any, can be used to determine the direction and 
the modulus of the velocity of the recoiling nuclei produced in the 
reaction. Fortunately, the 780~keV transition, located in the 
high-energy part of the level scheme (this is discussed below, see
sect.~\ref{level scheme}), displays only shifted components. This can be seen
in the triple gated spectrum drawn in figure~\ref{gate_587keV_TQC}. 
The 780~keV transition is shifted and broadened by the
Doppler effect, whereas the 727~keV ($2^+ \to 0^+$) transition 
is emitted by a fully stopped nucleus. 
\begin{figure}[!h]
\begin{center}
\includegraphics*[scale=0.25]{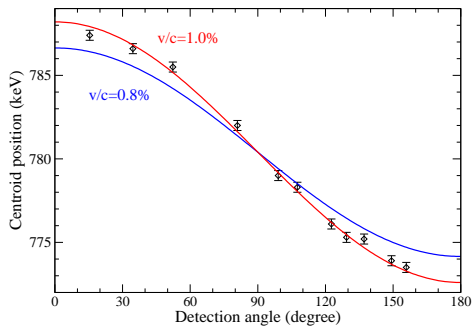}
\caption{(color online) Energy of the Doppler-shifted line around 780~keV as a 
function of the detection angle. The red curve is the best fit 
obtained for a relative velocity $\frac{v}{c}$=1\% and a transition
energy of 780.4~keV. The blue curve corresponds to the velocity of the 
$^{226}$Th compound nucleus, $\frac{v}{c}$=0.8\%.
\label{doppler_vitesse}  
}
\end{center}
\end{figure}
We have precisely measured the energy of the 780~keV transition 
as a function of the detector angle, the results are displayed in 
fig.~\ref{doppler_vitesse}. The symmetry around 90$^\circ$  
implies that the $^{212}$Po nuclei recoil along the
beam axis. The best fit gives $\frac{v}{c}$=1\% and a transition 
energy of 780.4~keV. 

The direction and the modulus of this velocity indicate that
the  $^{14}$C ejectiles are also emitted along the beam axis,
but in the backward direction\footnote{If the $^{14}$C ejectiles
were emitted forward, the velocity of $^{212}$Po would be smaller than
the velocity of $^{226}$Th, the compound nucleus of the 
$^{18}$O+$^{208}$Pb complete fusion, i.e. 0.8\%.}, 
which is in perfect agreement with
previous results. Many years ago, transfer reactions
induced by $^{16}$O on $^{208}$Pb have been studied~\cite{vi77} in the 
energy range of 77 to 102~MeV, the scattered particles being 
identified using Si telescopes and their angular distribution 
measured in interval of 5-10$^\circ$ over the angular range of 
30$^\circ$ to 
160$^\circ$ in the laboratory system. The results
are typical of a grazing collision, showing an energy dependence of
the maximum of the distribution, which moves from 60$^\circ$ for
102~MeV incident beam to more than 160$^\circ$ for 80~MeV. Some 
years later, using 
the transfer reaction $^{208}$Pb($^{18}$O, $^{14}$C) at bombarding
energies below 80~MeV~\cite{bo81}, the first excited states of 
$^{212}$Po have been identified from the coincidences
between backscattered ejectiles (within an angular range of 
151$^\circ$ $\le$ $\theta _{lab}$ $\le$ 166$^\circ$) and $\gamma$-rays.

\subsection{Level scheme of $^{212}$Po}\label{level scheme}
We fully agree with the results previously obtained on the level 
scheme of $^{212}$Po~\cite{po87,po03,ga05}, confirming 
both the identified $\gamma$-rays and their placement in the level 
scheme. Moreover we have 
assigned about 50 new $\gamma$-rays to $^{212}$Po, de-exciting 
35 new excited states, about ten of which are located above
2.92~MeV, the energy of its (18$^+$) long-lived state. 

\begin{figure*}[t!]
\begin{center}
\resizebox{0.88\textwidth}{!}{\includegraphics*{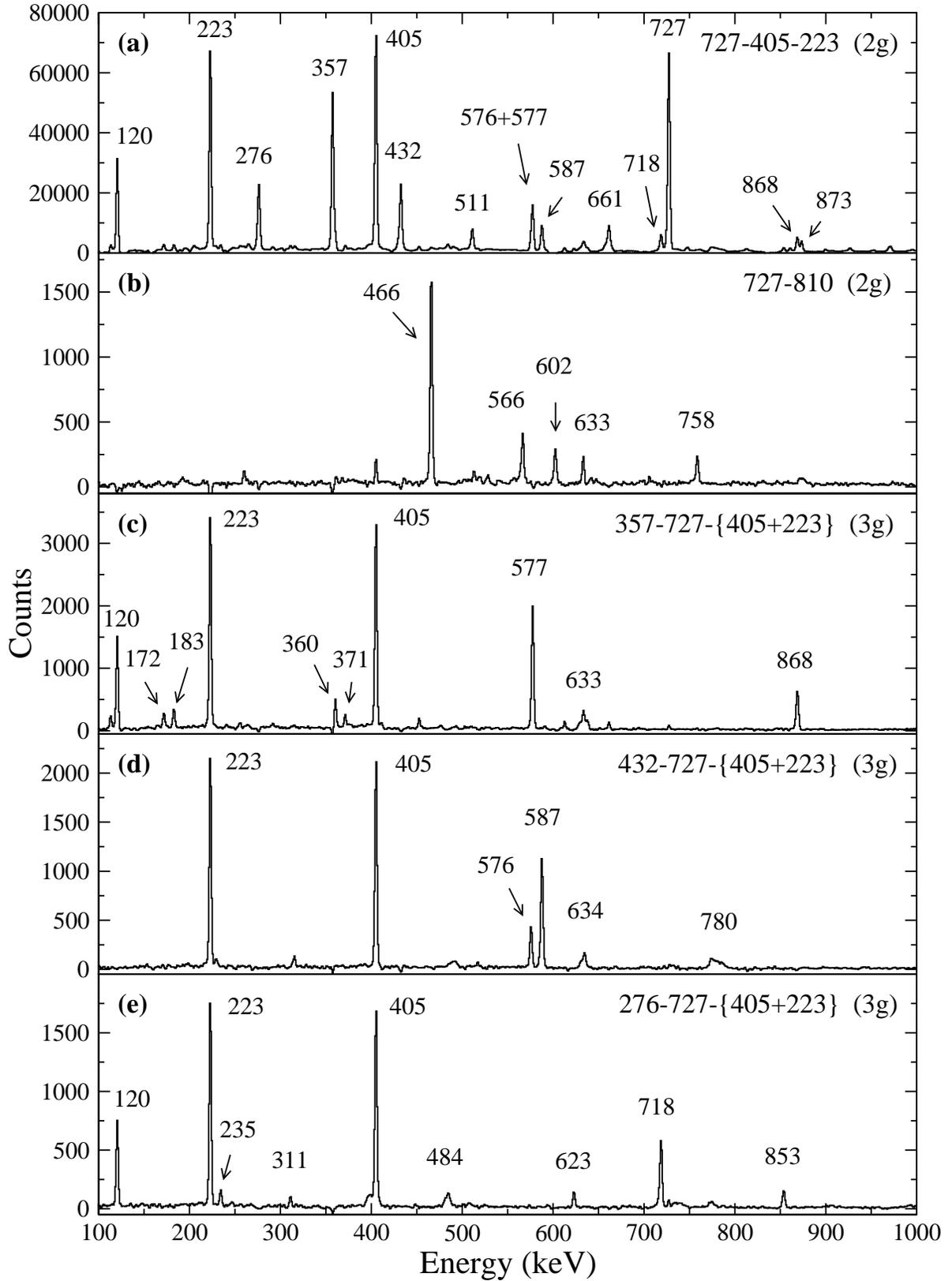}}
\caption{Background subtracted $^{212}$Po spectra built from the following $\gamma$-coincidence 
conditions: 
(a) 2 gates among 727, 405 and 223 keV; 
(b) 2 gates: 727 and 810 keV; 
(c) 3 gates: 357, 727 and (405 or 223) keV;
(d) 3 gates: 432, 727 and (405 or 223) keV;
(e) 3 gates: 276, 727 and (405 or 223) keV.
 }
\label{spectres}       
\end{center}
\end{figure*}

Typical coincidence spectra of $^{212}$Po are presented in 
figure~\ref{spectres}. 
By requiring two conditions among the three most intense lines 
(727, 405 and 223 keV), 
almost all transitions of $^{212}$Po can be seen, the most 
intense ones are labeled in figure~\ref{spectres}a. 
The 810~keV transition is located just above the 727~keV level
line and cannot be seen in this spectrum. The spectrum of $\gamma$-rays in
double coincidence with the 727~keV and 810~keV transitions (see 
figure~\ref{spectres}b) shows the
transitions belonging to the part of level scheme lying above 
the 810~keV transition.
The three last spectra displayed in 
figures \ref{spectres}c, \ref{spectres}d 
and \ref{spectres}e have been obtained by triple gating on three
transitions of $^{212}$Po, showing different parts 
of the level scheme which are well populated by the reaction 
used in this work. 
In the spectrum gated by two transitions of the yrast band and 
by the 432~keV line, one can even notice the Doppler-broadened 
780~keV transition, which has allowed us to extract the velocity 
of the recoiling nuclei (see section~\ref{kinematics}).

All the $\gamma$-rays observed in the reaction $^{208}$Pb($^{18}$O, $^{14}$C) are
reported in table~\ref{tableau_intensites}, including their relative
intensity, the value of their anisotropy, 
R$_{ADO}$=I$_{\gamma}$(39.3$^\circ$)/ I$_{\gamma}$(76.6$^\circ$) and 
their location in the level scheme. 
\begin{table*}
\caption[h]{\label{tableau_intensites} 
$\gamma$-ray transition energies (E$_\gamma$), relative $\gamma$ intensities 
(I$_\gamma$), angular distribution ratios ($R_{ADO}$), and level and spin 
assignments for the $^{212}$Po nucleus. 
}

\begin{center}
\begin{tabular}{|c|c|c|c|c|c|c||c|c|c|c|c|c|c|}
\hline
 E$_\gamma^{(a)}$ & I$_\gamma^{(b)}$  & $R_{ADO}$ & $E_i$ & $E_f$ & $I_i^\pi$ & $I_f^\pi$ 
& E$_\gamma^{(a)}$ & I$_\gamma^{(b)}$  & $R_{ADO}$ & $E_i$ & $E_f$ & $I_i^\pi$ & $I_f^\pi$ \\
  {\rm (keV)}      &        &               &     {\rm (keV)}  &     {\rm (keV)} &  &      
  &  {\rm (keV)}      &        &               &     {\rm (keV)} &     {\rm (keV)} & & \\
\hline
 69.2   & 6(3)   & 		& 2769.4 & 2700.4 & $13^- $  & $12^+$  		&566.3 & 14(3) & 1.32(14)& 2102.9& 1536.8& $5^{(-)}$ & $ 3^{(-)}$\\
 113.3 & 3(1)    & 		& 2882.8 & 2769.4 &  $ 14^+ $  & $13^- $ 	& 575.6& 10(3)   & 1.15(8) &2362.7 & 1786.9  & $6$ & $6^-$ \\  
 120.3 & 90(10) & 1.28(18) 	& 1474.9 & 1354.6 & $8^+ $  & $6^+$ 		& 577.1& 49(7)    & 0.82(2) &2409.1 & 1832.0  &  $11^- $  & $10^+$ \\ 
 157.2 &  4(2)   &		& 2102.9 & 1945.6 & $5^{(-)}$ &	$4^-$	        & 587.5& 29(5)     & 0.80(4) &2374.2 & 1786.9  & $7^{(-)}$ & $6^-$ \\
 171.7 & 5(2)    &  	     	& 2580.8 & 2409.1 & $ $  & $11^-$  		& 601.9& 9(3)     & 0.89(15)    &2604.4 & 2002.5 &$5$& $4^{(-)}$\\
 182.6 & 5(2)   & 1.37(16) 	& 2882.8 & 2700.4 &  $14^+ $  & $12^+$          &612.3$^\ast$& 27(5)& 1.24(27)&1744.3 & 1132.0 &  $4^-$  & $4^+$ \\  
 205.1 & 5(2)   &	  	& 2432.8 & 2227.7 &          & 7	       & 622.6& 3(1)     & 0.78(10) &2374.2 & 1751.0& $7^{(-)}$  & $8^-$\\ 
 222.6 & 530(30) & 1.21(3) 	& 1354.6 & 1132.0 & $6^+ $  & $4^+$ 		&  633.2& 9(3)      & 0.82(13)&2170.0 & 1536.8  & $ $  & $3^{(-)}$  \\
 229.8 & 3(1)    &		& 2604.3 & 2374.2 &   & $7^{(-)}$ 		&  633.3$^\ast$&15(3)&1.20(16)&2465.3&1832.0&$10^-$& $10^+$\\
 234.6 & 4(2)    & 1.16(13)     & 1985.7 & 1751.0 & $8^{(-)}$  & $8^-$ 		& 633.6& 9(3) &	&2420.5 &1786.9  &&  $6^-$ \\ 
 249.7 & 3(1)    & 	 	& 2235.5 & 1985.7 &	& $8^{(-)}$			& 637.3 & 2(1) &		&2469.6 &1832.0  &$9^{(-)}$& $10^+$ \\
 255.2 &  2(1)   &		& 3035.3 & 2780.2 & $ $  & $10$			&  661.3$^\ast$&51(7)&1.19(9) &2015.9 & 1354.6  & $6^-$  & $6^+$   \\
 259.6 & 5(2)    &		& 2362.7 & 2102.9 & $6$  & $5^{(-)}$ 	        &  718.4& 20(5) 	&0.76(4) &2469.6 & 1751.0  & $9^{(-)}$  & $8^-$ \\
 264.7 & 7(2)   & 0.99(9)	& 2280.7 & 2015.9 && $6^-$			& 727.1   &  -- 	&1.14(2) &727.1 & 0.0 &  $2^+ $  & $0^+$  \\ 
 276.1$^\ast$&58(10)&1.28(7)   & 1751.0 & 1474.9 & $8^-$  & $8^+ $ 	 	&  740.2$^\ast$&7(3) &		&3209.8& 2469.6  &  $ $  & $9^{(-)}$ \\
 291.4 & 3(2)    &		& 3174.2 & 2882.8 & $(15)$  & $14^+$  		& 748.1& 6(3) &		&2102.9 & 1354.6  & $5^{(-)}$  & $6^+ $ \\
 310.8  & 3(2)   &		& 2780.2 & 2469.6 &  $10$  & $9^{(-)}$ 		& 757.2$^\ast$& 8(3) &		&2860.1 & 2102.9  && $5^{(-)}$ \\ 
 315.5 & 3(2)    &		& 2102.9 & 1786.9 & $5^{(-)}$ &	$6^-$		& 758.4& 10(3) & 1.15(11) 	&2295.2 & 1536.8  & $ $  & $3^{(-)}$  \\ 
 357.1 & 170(20) &1.26(3)	& 1832.0 & 1474.9 & $ 10^+ $  & $8^+ $		& 774 & 2(1)  & 		&2525.0 & 1751.0  & $ $  & $8^-$  \\ 
 358.5 & 11(3)   &  	 	& 2102.9 & 1744.3 & $5^{(-)}$ & $4^-$	  	& 780.4$^\ast$& 23(5) & 1.34(9)  &3154.7 & 2374.2  & $7^{(+)}$& $7^{(-)}$ \\ 
 358.6 & 11(3)   & 0.85(10) 	& 2374.2 & 2015.9 & $7^{(-)}$  & $6^-$		& 809.7&  76(10) & 0.83(6) 	&1536.8 & 727.1 &  $ 3^{(-)}$  & $2^+$  \\
 360.2 & 10(3)   & 1.26(15)	& 2769.4 & 2409.1 & $ 13^- $  & $11^- $		& 813.6$^\ast$&  34(5) & 1.13(8) &1945.6 & 1132.0  & $4^-$  & $4^+$ \\ 
 371.0 & 3(1)    &	 	& 2780.2 & 2409.1 & $10$     &	$11^- $	 	& 853.4& 7(2) & 0.98(20)     &2604.3 & 1751.0  &   & $8^-$ \\ 
 397.7$^\ast$&9(3)&	  	 & 2867.3 & 2469.6 &  $ $  & $9^{(-)}$  	& 860.3& 5(2) & 0.74(15)	&2335.1 & 1474.9  & 9 & $8^+$\\ 
 404.9 &  924(25)& 1.19(2) 	& 1132.0 & 727.1  &  $4^+ $  & $2^+$  		&   868.4& 19(4) & 1.18(6) 	&2700.4 & 1832.0  &  $12^+ $  & $10^+$ \\
 405  & 6(3)     &	  	& 1536.8 & 1132.0 & $3^{(-)} $  & $4^+$  	&  873.1& 14(3) & 0.84(5) 	&2227.7 & 1354.6  & $7$  & $6^+ $ \\
 406.6$^\ast$ &2(1) &	  	 & 3010.9&  2604.3   &&         		& 875.0 & 4(2) &		&2877.5 & 2002.5  &&  $4^{(-)}$  \\ 
 410.7 & 2(1)    & 1.08(16) 	& 3111.1 & 2700.4  &  $ $  & $12^+$		& 899.0 & 4(2) &		&2374.2 & 1474.9  & $7^{(-)}$  & $6^+ $  \\ 
 432.3$^\ast$&75(10)&1.22(6)	 & 1786.9 & 1354.6  & $6^-$  & $6^+$	 	 & 926.2 & 6(3) & 0.67(14) 	&2280.7 & 1354.6  & $ $  & $6^+ $ \\ 
 452.3 & 4(2)    &		& 2861.4 & 2409.1  && $11^- $			& 953.1& 8(3) &1.13(29)	&2085.1 & 1132.0  & $ $  & $4^+$ \\ 
 465.7 & 50(5)   &0.89(9)	& 2002.5 & 1536.8  & $4^{(-)}$  & $3^{(-)}$ 	 & 968.9& 14(3) &0.86(10)	&2100.9 & 1132.0  & $5$  & $4^+$\\ 
 474.9 & 3(1)    &		& 2940.2 & 2465.3  &		& $10^-$ 	&  971.1& 26(5) &0.79(7)	&2102.9 & 1132.0  & $5^{(-)}$  & $4^+$\\
 483.7 & 4(2)    &  		& 2469.6 & 1985.7  &  $9^{(-)}$  & $8^{(-)}$ 		&  994.9& 4(2) &  		&2469.6 & 1474.9  &  $9^{(-)}$  & $8^+$ \\
 484.5 & 7(3)   & 		& 2235.5 & 1751.0  &   & $8^-$     		&  1005$^\ast$& 6(3) &  	&2837 & 1832.0  &    & $10^+$ \\
 490.2$^\ast$& 5(2)& 1.26(15)	 & 2864.5 & 2374.2  &$7^{(+)}$&  $7^{(-)}$ 	&  1020& 8(2) &   0.76(7)	&2374.2 & 1354.6  & $7^{(-)}$  & $6^+ $ \\
 492.8 & 2(1)    &		& 3193.2 & 2700.4  &   & $12^+ $		&  1049$^\ast$& 9(3)& 1.03(17) & 2881 & 1832.0 &  & $10^+$\\
 502.9 & 3(1)    &  		& 2335.1 & 1832.0  & 9 & $10^+$ 		&  1172.7& 2(1) &  		&3004.7 & 1832.0  &    & $10^+$ \\
 510.9 & 26(5)   &1.10(9)	& 1985.7 & 1474.9  & $8^{(-)}$  & $8^+ $	& 1371$^\ast$& 6(3) &1.16(20)&3203& 1832.0 &  & $10^+$ \\
 518.4 & 3(1)    &		& 2880.9 & 2362.7  && $6$ 			&  1620$^\ast$& 6(3) &  	&2975 & 1354.6  &   & $6^+ $ \\
563.8$^\ast$& 10(3)   &          & 2666.7 &  2102.9 &  & $  5^{(-)}$ 		&  1669$^\ast$ & 5(2) &  	&3024 & 1354.6&   & $6^+ $ \\ 
\hline
\end{tabular}
\end{center}
$^{(a)}$ 
Uncertainties in transition energies are typically between 0.1 and 0.5~keV. 
Transitions showing Doppler shift and broadening are marked by an asterisk, 
their energy uncertainty can be larger (up to 2~keV).\\
$^{(b)}$
Intensities measured in this experiment (i.e. with the requirement that a minimum of
three unsuppressed Ge detectors fired in prompt coincidence) are normalized to the sum of the values of the two transitions 
populating the 2$^+_1$ state, I$_\gamma$(404.9~keV) + I$_\gamma$(809.7~keV)~=~1000.\\
\end{table*}
Results of the angular distributions and
of angular correlations for the most intense $\gamma$-rays are given in 
tables~\ref{distri_angul} and \ref{angular_correl}. 
Values of conversion coefficients for the low-energy transitions extracted
from total intensity balances done on some selected gated spectra are 
given in table~\ref{ICC}, in comparison with theoretical
values~\cite{ki08}.
\begin{table}[!h]
\begin{center}
\caption{Angular distribution coefficient (a$_2$) and multipole order of the most 
intense $\gamma$-rays of $^{212}$Po. The transitions
marked with an asterisk are discussed in 
sects.~\ref{conversion}~and~\ref{lifetime}. 
}
\label{distri_angul}
\begin{tabular}{|cc|cc|}
\hline
E$_\gamma$ & a$_2^{(a)}$ & multipole order & spin sequence$^{(b)}$\\
\hline
727.1  	&$+0.16(4)$   & $\Delta$I=2 quadrupole	&  $2^+ \to 0^+$	\\
404.9  	&$+0.20(4)$   & $\Delta$I=2 quadrupole	&  $4^+ \to 2^+$	\\
222.6  	&$+0.18(3)$   & $\Delta$I=2 quadrupole	&  $6^+ \to 4^+$	\\
357.1  	&$+0.24(4)$  & $\Delta$I=2 quadrupole	& $10^+ \to 8^+$	\\
577.1  	&$-0.25(5)$   & $\Delta$I=1 dipole	& $11^- \to 10^+$	\\
868.4  	&$+0.18(6)$   & $\Delta$I=2 quadrupole	& $12^+ \to 10^+$	\\
  &   & & \\
276.1*	&$+0.29(6)$   & $\Delta$I=0 dipole	& $8^- \to 8^+$	\\
432.3*	&$+0.27(7)$   & $\Delta$I=0 dipole	& $6^- \to 6^+$	\\
465.7	&$-0.5(1)$   & $\Delta$I=1 dipole	& $4^{(-)} \to 3^{(-)}$	\\
575.6	&$+0.3(1)$   & $\Delta$I=0 dipole	& $6 \to 6^-$	\\
587.5	&$-0.23(8)$   & $\Delta$I=1 dipole	& $7^{(-)} \to 6^-$	\\
661.3*	&$+0.26(9)$   & $\Delta$I=0 dipole	& $6^- \to 6^+$	\\
718.4	&$-0.29(5)$   & $\Delta$I=1 dipole	& $9^{(-)} \to 8^-$	\\
1020	&$-0.28(8)$   & $\Delta$I=1 dipole	& $7^{(-)} \to 6^+$	\\
\hline
\end{tabular}
\end{center}
$^{(a)}$ The number in parenthesis is the error in the last digit.\\
$^{(b)}$  see text.
\end{table}

\begin{table}[!h]
\begin{center}
\caption{Coincidence rates between $\gamma$-rays of $^{212}$Po 
as a function of their relative angle of detection, normalized to 
the ones obtained around 75$^\circ$.
}
\label{angular_correl}
\begin{tabular}{|c|ccc|}
\hline
E$_\gamma$-E$_\gamma$ & R(22$^\circ$) & R(46$^\circ$) & R(75$^\circ$)\\
\hline
 405~-~727  &  1.10(4) & 1.06(2)	&1.00(1)	\\
223~-~405  &  1.10(3) & 1.05(2)	&1.00(1)	\\
577~-~357  &  0.89(5) & 0.97(4)	&1.00(3)	\\
  &   & & \\
223~-~276  &  1.12(6) & 1.09(4)	&1.00(2)	\\
223~-~432  &  1.15(6) & 1.12(4)	&1.00(3)	\\
223~-~661  &  1.22(7) & 1.08(5)	&1.00(3)	\\
\hline
\end{tabular}
\end{center}
\end{table}

\begin{table}[!h]
\begin{center}
\caption{Conversion coefficients for the low-energy
transitions of $^{212}$Po extracted from intensity balances done on 
gated spectra, the theoretical values come from the $BrIcc$ 
database~\cite{ki08}. The transitions
marked with an asterisk are discussed in 
sects.~\ref{conversion}~and~\ref{lifetime}.
}
\label{ICC}
\begin{tabular}{|ccccccc|}
\hline
 E & &$\alpha$ & $\alpha$ & $\alpha$ & $\alpha$& adopted\\
 (keV)& &(exp) & (E1)&(E2) &(M1)&\\
\hline
69.2  & tot 	& 0.2(1)    & 0.26   	& 39.2	& 6.2	&E1\\
113.3 & tot 	& $<$ 3.0    & 0.33   	& 4.26 	 & 7.77	 &E1\\
120.3& tot 	&  3.2(5)   & 0.29   	& 3.32	& 6.60  &E2\\ 
	 & K	&  	    &  	        &0.41	 &  	&\\
182.6 & tot 	& 0.6(3)    & 0.10   	& 0.65	& 2.00 	&E2\\
222.6& tot 	&  0.33(5)  & 0.063  	& 0.33	& 1.15	&E2\\ 
	 & K	&  	    &  	        &0.13	 &  	&\\
276.1*& tot 	&  0.37(7)  & 0.038 	&0.162	& 0.635	&E1$^{(a)}$\\  
	& K	& $\sim$ 0.4 & 0.031	 &0.081	 & 0.516 &\\ 
432.3*& tot 	&  0.13(3)  & 0.014	&0.046	& 0.188	&E1$^{(a)}$\\ 
	 & K	& $\sim$ 0.1 & 0.011	 &0.030	 & 0.153 &\\
\hline
\end{tabular}
\end{center}
$^{(a)}$ see sect.~\ref{conversion}
\end{table}

The main feature of the level scheme of $^{212}$Po obtained in 
the present work is the rather low multiplicity of the $\gamma$ 
cascades, meaning that the $\alpha$ transfer reaction induced by 
$^{18}$O at a bombarding energy close
to the Coulomb barrier does not provide so much angular momenta. 
\begin{figure}[!h]
\begin{center}
\resizebox{0.45\textwidth}{!}{\includegraphics*{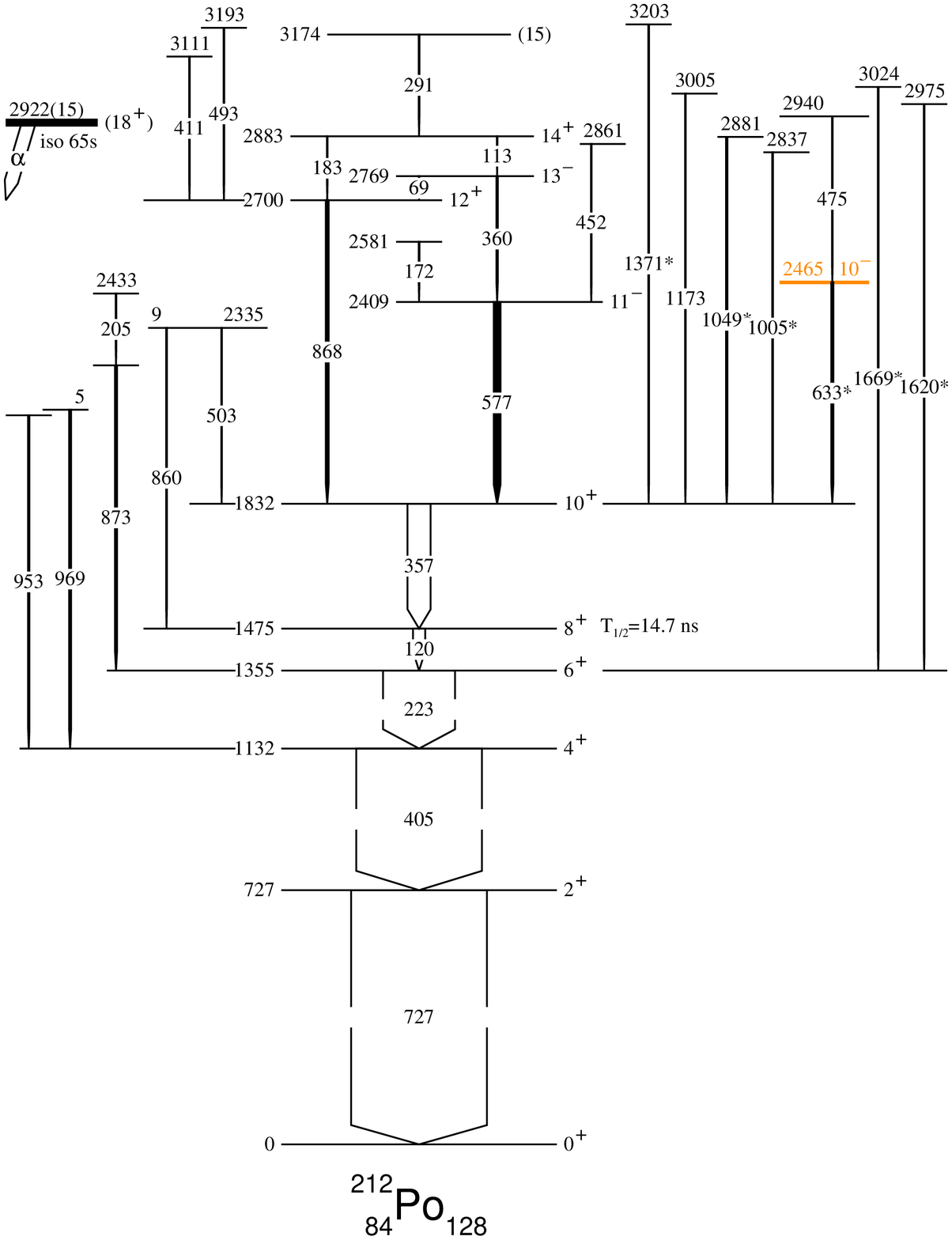}}
\caption{(color online) First part of the level scheme of $^{212}$Po 
determined in this work, 
showing the yrast part and levels only linked to yrast states. The 
long-lived isomeric state at 2922(15)~keV excitation energy~\cite{NNDC}, 
a pure $\alpha$-emitter not observed in the present work, is drawn for 
the sake of completeness. The half-life of the 8$^+$ yrast state is
14.7(3)~ns (see text).
The width of the arrows is representative of the intensity of the
$\gamma$-rays. The transitions
marked with an asterisk exhibit Doppler shifts (see sect.~\ref{lifetime}).
The colored state is also displayed in fig.~\ref{States_212Po}.}
\label{schema_popo1}       
\end{center}
\end{figure}
The new states which have been identified are lying in a medium range of 
excitation energy (2.5~MeV-3.2~MeV) and angular momentum (6$\hbar$-8$\hbar$).
For the purpose of clarity, the level scheme has
been split in two almost independent parts which are displayed in 
figures~\ref{schema_popo1} and \ref{schema_popotwo}.
\begin{figure*}[!t]
\begin{center}
\includegraphics*[angle=90,scale=0.60]{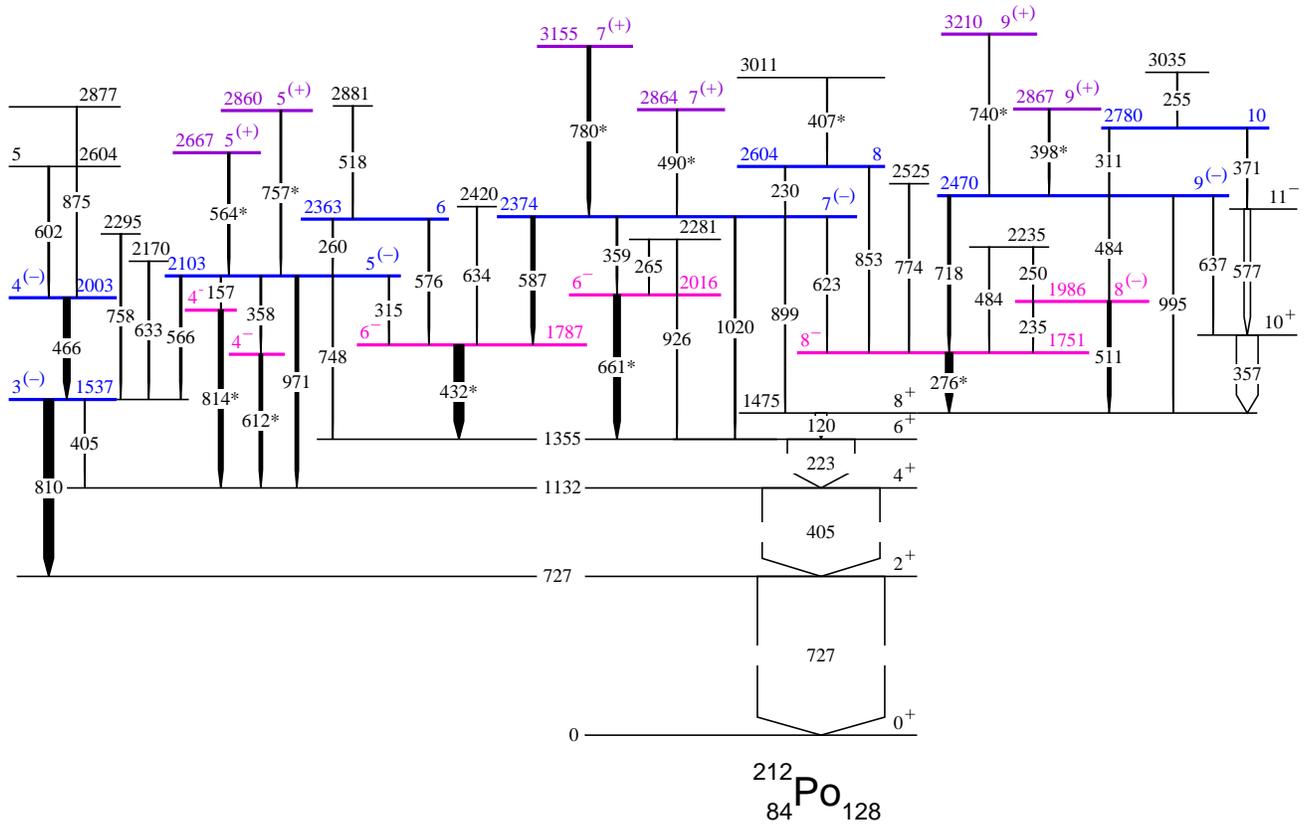}
\caption{(color online) Second part of the level scheme of $^{212}$Po determined in this work, 
showing several groups of excited states linked together and to the yrast 
states. The width of the arrows is representative 
of the intensity of the $\gamma$-rays. The transitions
marked with an asterisk exhibit Doppler shifts (see sect.~\ref{lifetime}).
The colored states are also displayed in fig.~\ref{States_212Po}.
}
\label{schema_popotwo}       
\end{center}
\end{figure*}

The first part of the level scheme (see fig.~\ref{schema_popo1}) shows the 
yrast states and  the levels only linked to yrast states. We have extended the
yrast cascade by only one transition, 291~keV, whereas many new
$\gamma$-rays have been found to populate directly the lower-spin yrast
states, some of them having a high energy. 
It is worth pointing out that both the 1620~keV and 1669~keV transitions 
are seen in coincidence with the 727~keV, 405~keV and 
223~keV transitions, while their possible coincidence with the 120~keV line
could be too lowered because of its very low $\gamma$ intensity. 
We have chosen to place the two high-energy transitions
directly above the 1335~keV state, but their location above the 
1475~keV state cannot be excluded.
The spin and parity values of all the yrast states are now determined 
unambiguously up to the 2883~keV level with I$^\pi$=14$^+$, on the basis of 
our results for angular distributions (table~\ref{distri_angul}), 
angular correlations (table~\ref{angular_correl}), ADO ratios 
(table~\ref{tableau_intensites}),
as well as the internal conversion electron coefficients of the
low-energy transitions (table~\ref{ICC}). 
Because of the E1 multipolarity of the 69~keV and the 113~keV transitions,
the negative parity of the 2409~keV and the 2769~keV states is now clearly
established. 

The timing information from the germanium detectors have been used to measure
the half-life of the 8$^+$ yrast state. The procedures were checked using 
the delayed coincidences of isomeric states of various nuclei
produced in this experiment, as described in ref.~\cite{as06}. The time 
distribution between the emission of $\gamma$-rays located above the 1475~keV
level and those below it gives T$_{1/2}$~=~14(1)~ns. This is lower than the
value obtained in ref.~\cite{po87}, which has been adopted in the ENSDF data
base~\cite{NNDC}. On this other hand, our value is in better agreement with the 
two previous results~\cite{li78,bo81}, T$_{1/2}$~=~14.2(24)~ns and 
T$_{1/2}$~=~14.7(3)~ns, respectively. Then, we select the most accurate
value, as shown in fig.~\ref{schema_popo1}. 

The second part of the level scheme (see fig.~\ref{schema_popotwo}) shows
several groups of excited states linked together and to the  
yrast states. 
Some levels display several decay paths which
limit the spin values which can be assigned to the involved levels. 
For instance, the angular distribution results for the 810~keV, 
971~keV and 1020~keV 
transitions (see tables~\ref{tableau_intensites} and \ref{distri_angul})
indicate that they are dipole 
transitions linking states with $\Delta I = 1$. Thus the spin
values of the 1537~keV, 2103~keV and 2374~keV states are 3, 5, 
and 7, respectively, because of their other decay observed 
towards the yrast state having one unit more angular momentum 
(see fig.~\ref{schema_popotwo}). Knowing that the 587~keV,
359~keV, and 623~keV transitions are dipole 
transitions linking states with $\Delta I = 1$, the spin values of
the 1787~keV, 2016~keV states are {\it even}.
As for the 432~keV transition linking two levels having an
even-$I$ value, it could 
be assigned quadrupole transition with 
$\Delta I = 2$, as its $a_2$ angular coefficient is
+0.27(7) (see table~\ref{distri_angul}), or its coincidence
rate with the 223~keV transition is maximum at 22$^\circ$ (see
table~\ref{angular_correl}).
Nevertheless the $I^\pi$ value of
the 1787~keV state cannot be 8$^+$, because of its link to 
the 2103~keV state ($I$~=~5) by means of the 315~keV transition.  
It is worth recalling that a {\it pure} dipole $\Delta I = 0$ transition
has almost identical angular distribution/angular correlations 
as a quadrupole $\Delta I = 2$ one. Thus 
the spin value of the 1787~keV state is fixed to 6. 
Four other $\gamma$-rays drawn in fig.~\ref{schema_popotwo} 
(276~keV, 612~keV, 661~keV, and 814~keV) share the same
features as the 432~keV transition. They exhibit Doppler shifts, 
they are located just above an yrast state, their angular properties 
indicate that they can be either quadrupole transitions linking states with 
$\Delta I = 2$  or {\it pure} dipole transitions linking states with 
$\Delta I = 0$. Thus we adopt the same conclusion for all 
these transitions, they are dipole transitions linking states with 
$\Delta I = 0$, in agreement with the other decay paths of the states
located above them. The negative parity of the 1744~keV, 1751~keV, 1787~keV, 
1946~keV, and 2016~keV states 
will be discussed in the next section.  
    
Due to coincidence relationships and/or spin values, we propose 
two different states at 2604~keV, one deexcites towards the 2003~keV 
state by means of the 602~keV transition, while the second one
is linked to the 2374 keV state and to the
1751~keV one (see fig.~\ref{schema_popotwo}).  
Secondly, two transitions at 968.9-971.1~keV deexcite 
two states close in energy (2100.9-2102.9~keV). Both members of the 
doublet, being clearly seen in coincidence with the two first yrast 
transitions, are located just above the 1132.0~keV level 
(see figs.~\ref{schema_popo1} and \ref{schema_popotwo}). Whereas there 
is no other transition in coincidence with the lower-energy member, 
the higher-energy member exhibits the same coincidence
relationships as the 566~keV transition (see fig.~\ref{schema_popotwo}).
Lastly, we have assumed that there is only one level at 2335~keV 
(deexcited by 861 and 503~keV transitions, see 
fig.~\ref{schema_popo1}), as well as only one level at 2281~keV 
(deexcited by the 265~keV and 926~keV transitions, see
fig.~\ref{schema_popotwo}) and one level at 2235~keV 
(deexcited by the 250~keV and 484~keV transitions) 
 but we could not strengthen these three 
hypotheses because of lack of coincidence relationships. 

\subsection{Anomalous conversion coefficients of the 276~keV and 
432~keV transitions}\label{conversion}
We have extracted the internal conversion electron 
coefficients of some transitions of $^{212}$Po by analyzing 
the relative intensities of transitions in cascade. 
The intensity 
imbalances of the 120.3~keV and the 222.6~keV transitions 
measured in spectra in double coincidence with at least one
$\gamma$-ray located above them lead to 
$\alpha_{tot}$(120~keV)=3.3(4) and 
$\alpha_{tot}$(223~keV)=0.33(4), in good agreement with the
theoretical values for E2 multipolarity~\cite{ki08}.
In addition (see table~\ref{ICC}), we have found 
$\alpha_{tot}$(276~keV)=0.37(7) and 
$\alpha_{tot}$(432~keV)=0.13(3). 

\begin{figure}[!h]
\begin{center}
\resizebox{0.4\textwidth}{!}{\includegraphics*{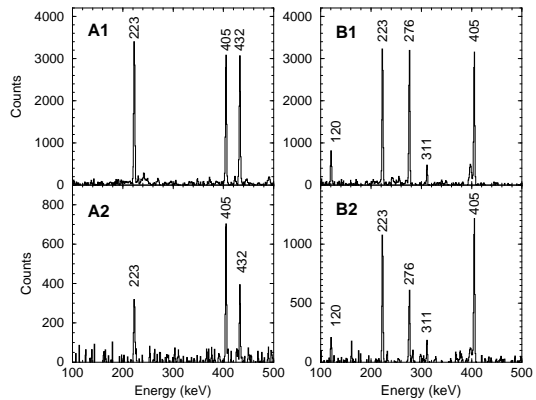}}
\caption{{\bf Left}: spectra of $\gamma$-rays in double coincidence with 
the 587~keV and 727~keV transitions (A1) and with the 587~keV and 79~keV 
(K$_{\alpha _1}$ X-ray of Po) transitions (A2). {\bf Right}: 
spectra of $\gamma$-rays in double coincidence with the
718~keV and 727~keV transitions (B1) and with the 718~keV and 79~keV 
transitions (B2).}
\label{coinc79}       
\end{center}
\end{figure}

The vacancies in the K electron shell due to the emission of an 
internal conversion 
electron lead to X-ray emission and conversely the rate 
of X-rays emitted in any cascade is correlated to the $\alpha_K$
values of the involved transitions. We have exploited this 
feature to measure some $\alpha_K$ coefficients.
For instance, in the spectra in double coincidence with the
79~keV K$_{\alpha _1}$ X-ray of Po and the 577~keV (or 357~keV)
transition, the loss in intensity of the 120~keV transition is 
three times higher than the one  of the 223~keV transition, in 
agreement with the ratio of their $\alpha_{K}$ 
coefficient~\cite{ki08}. 
In the spectrum in double coincidence with the
587~keV $\gamma$-ray and the 79~keV X-ray, the 432~keV 
transition  exhibits the same drop in intensity as
the 223~keV one (compare the A1 and A2 spectra of fig.~\ref{coinc79}). 
Thus we obtain
$\alpha_{K}$(432~keV)~$\approx$~$\alpha_{K}$(223~keV, E2)(=~0.13).
In the same way (see the B1 and B2 spectra of fig.~\ref{coinc79}), 
we get
$\alpha_{K}$(276~keV)~$\approx$~$\alpha_{K}$(120~keV, E2)(=~0.41).

The multipolarities of the 276~keV and 432~keV transitions derived 
from their $\alpha_{tot}$ values could be M1+E2, implying a mixing 
$\delta^2 \sim 1$, $i.e.$  50\% M1 + 50\% E2. However, the occurrence 
of such a very large E2 component leads to major discrepancies:
\begin{itemize}
\item The $a_2$ angular coefficient computed for such mixed
multipolarities is no longer in agreement with the measured values, 
as the quadrupole component leads to a large decrease of the $a_2$  
coefficient (for $\delta$=+1) and even to a quasi-isotropic 
emission (for $\delta$=-1).  
\item The B(E2) reduced transition probability computed
from the lifetime of the 1751~keV state  (given in the next section)
largely exceeds the recommended upper limit (RUL) for $\gamma$-ray
strengths: B(E2, 276~keV) $\sim 5 \times 10^3~W.u.$, 
that is larger than what is obtained for E2 transitions {\it inside} 
the superdeformed bands in the Hg-Pb region, 
B(E2, SD) $\sim 2 \times 10^3~W.u.$. 
\end{itemize}
It is noteworthy that the E2 component of the 432~keV transition would
lead to B(E2) $\sim 6 \times 10^2~W.u.$. Such a large value would be 
the sign of a quadrupole deformation, implying a rotational 
behaviour, contrary to what is observed (the 120~keV transition
populating the same yrast state as the 432~keV transition has 
B(E2) $\sim 10~W.u.$, see fig.~\ref{BE2_PoPb}).

As well, the E1+M2 multipolarity that could account for the 
two $\alpha_{tot}$ values ($\sim$ 15--20\% M2) cannot be 
adopted because of the tremendous values of the B(M2) reduced 
transition probabilities calculated using the lifetimes of the 
states (given in the next section).

Thus we conclude that these transitions must have a pure E1 
multipolarity, with {\it anomalous} conversion coefficients.
In the theoretical calculations of the conversion 
coefficients such as those of ref.~\cite{ki08}, the nucleus is 
assumed to be a homogeneously charged {\it sphere}, i.e. 
without any electromagnetic moment, and the electron remains 
outside the nucleus. Nevertheless, it is well known that the 
atomic electrons can penetrate within the nuclear charge 
and current distributions, giving additional
nuclear matrix elements into the expression for the rate of
internal-conversion-electron ejection~\cite{ch56}. Then the
conversion coefficients may have {\it anomalous} values depending 
on nuclear structure~\cite{gr58,ni58}.
Furthermore, the nuclear electromagnetic moments add
several coupling terms to the Dirac equation of the
electron~\cite{pa75}.
Altogether, these effects should largely shift the values 
of the conversion coefficients of the 276~keV 
and 432~keV transitions since they are emitted by peculiar
states having a very large dipole moment, as outlined later 
(see sect.~\ref{discuss_cluster}).

As a result, the 1751~keV and  1787~keV states have a negative 
parity and we adopt the same conclusion for the 1744~keV, 
1946~keV, and 2016~keV states since their decaying $\gamma$-rays share
the same features as the 276~keV and 432~keV transitions, as above
mentioned. The negative parity of these five states accounts for 
their unique decay to the yrast state having the same $I$ 
value\footnote{In case of
positive parity, we would have to understand why these five states 
do not decay to the $(I-2)$ yrast states. For instance, the E2 decay 
of the 1787~keV state to the 4$^+$ yrast state at 1132~keV would be 
enhanced because of its energy by a factor of 8 as compared to the 
one to the 6$^+$ yrast state.}.

\subsection{Lifetime measurements}\label{lifetime}

As introduced previously, several $\gamma$-rays in $^{212}$Po exhibit 
Doppler shifts. A total of 18 transitions have been found to display 
this feature (they are marked with an asterisk in 
table~\ref{tableau_intensites}), implying that 18 excited states 
do have shorter lifetimes than the stopping time 
of $^{212}$Po in the lead target. 

Seven lines -- namely the 276~keV, 432~keV, 612~keV, 633~keV, 661~keV, 
780~keV and 814~keV transitions -- de-exciting some of those states, 
could be analyzed to get the lifetime using the Doppler-shift attenuation
method (DSAM), whereas in two cases (490~keV and 1049~keV transitions), 
only a limit of the value could be obtained. As for the 9 last cases, the
statistics was too low to perform any analysis  and the lifetimes for the
corresponding levels are assumed to be smaller than 1.4~ps, the mean value of
the stopping time of the $^{212}$Po recoils (reported below).

The kinematics of the
quasi-elastic transfer reaction was taken into account and fine tuned
using the line-shape of the 780~keV transition which displays 
predominantly a shifted part. The best results were obtained under the
assumption that the $^{212}$Po nuclei recoil in the forward direction
within a cone with an opening angle of $\pm 35^\circ$. 
The electron stopping power 
for $^{212}$Po ions in Pb was interpolated using the semi-empirical tables 
of Northcliffe and Schilling~\cite{Northcliffe_70} and
corrected  to take into account atomic structure 
effects~\cite{Ziegler_74,Ziegler_85_1}. 
The nuclear stopping power, which is
due to the interaction with the atoms of the medium as a whole,
was taken into account according to the LSS theory~\cite{LSS}
and the parameterization of the universal scattering function for a 
Thomas-Fermi potential given in Ref.~\cite{Currie_69}. To correct for
the effect of microchanneling in the stopping medium, the nuclear
stopping power was reduced by a factor $f_n$ = 0.7 
(cf. Refs.~\cite{Ziegler_85_1,Keinonen_84} for more details).
According to the calculation, the distribution 
of the  stopping times of the $^{212}$Po recoils is characterized by
a mean value of 1.35~ps and a variance of about 0.37$^2$~ps$^2$. 

The line-shapes of the  transitions to be analyzed were obtained 
using the coincidence matrices by setting
gates on fully stopped $\gamma$-ray peaks belonging to transitions 
depopulating levels lying below the level of interest. The response
functions of the cluster detectors in our experiment was characterized
by small asymmetry towards lower energies which was taken into account too. 
Further, for the line-shape analysis, the natural time-dependent 
functional form of the population $n_i(t)$ of the investigated level 
was used, namely,
\begin{equation}
n_i(t) = \sum_{k\geq i}C_{ik}exp(-\lambda_k t)
\end{equation}
which represents a superposition of exponentials with coefficients $C_{ik}$
determined by the decay constants $\lambda_k$ of the levels participating
in the cascade and by the branching ratios at every level. 
In the general case,
a fitting procedure which aims to determine simultaneously many decay
constants is hindered by the necessity to find a multiparameter solution,
including some parameters describing the non-observed or side-feeding.
In the present case, however, it was established for all levels 
depopulated by Doppler-shifted transitions that their observed
discrete feeding is slow, and responsible only for part of the intensity
of the unshifted fraction of the line-shape. The rest of the feeding is
very fast, corresponding to (nearly) direct population of the levels of
interest. This feature significantly simplifies the analysis, since
the intensity  balance at every level unambiguously fixes the areas of
the ``slow'' and ``fast'' fractions of the lineshape. Then, the only
remaining parameter to be determined is the decay constant of the level
of interest or its lifetime ($\tau = 1/\lambda$).

\begin{figure}[!h]
\begin{center}
\resizebox{0.40\textwidth}{!}{\includegraphics*{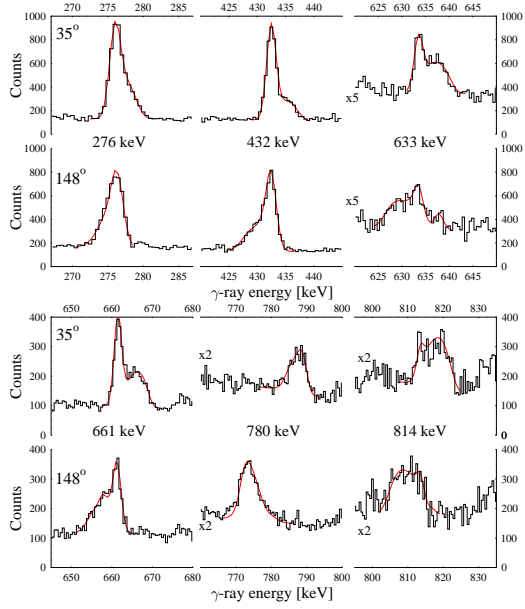}}
\caption[]{\label{line-shapes} (color online) Examples of line-shape analysis
at the forward angle of 35$^\circ$ and the backward angle of 148$^\circ$.}
\end{center}
\end{figure}
Examples of line-shape analysis at the forward angle of 35$^\circ$ 
and the backward angle of 148$^\circ$
are displayed in figure~\ref{line-shapes}, and
all the results are summarized in tables~\ref{lifetimes} and
~\ref{lifetimes_limit}.
\begin{table}[!h]
\begin{center}
\caption[h]{\label{lifetimes} Lifetimes measured in the present work with
their uncertainties. The decomposition of the feeding intensity into 
slow and fast fractions is also indicated.}
\begin{tabular}{|c|cc|c|c|}
\hline
State &  \multicolumn{2}{c|}{Feeding$^{(a)}$} & Decay & $\tau$$^{(a)}$ \\
    (keV)           &  Slow$[\%]$  & Fast$[\%]$   &  E$_\gamma$(keV) & (ps)\\
\hline
    1744  & 	 33(7)   &	67(7)    &   612.3  &  0.48(15)  \\
    1751  &      68(4)    &     32(4)    &   276.1  &  0.48(20) \\
    1787  &      68(3)    &     32(2)    &   432.3  &  0.45(8)  \\
    1946  &      13(4)    &     87(4)    &   813.6  &  0.47(15) \\
    2016  &      44(3)    &     56(3)    &   661.3  &  0.49(16) \\
    2465 &       30(4)   &	70(4)    &   633.3  &  0.61(16) \\
    3155  &    $\approx$ 0  &$\approx$100  &   780.4  &  0.12(6)  \\
\hline
\end{tabular}
\end{center}
$^{(a)}$ the number in parenthesis is the error in the last digit.
\end{table}
\begin{table}[!h]
\begin{center}
\caption[h]{\label{lifetimes_limit} Limit values of the lifetimes 
derived in the present work.}
\begin{tabular}{|c|c|c||c|c|c|}
\hline
State & Decay & $\tau$$^{(a)}$ & State & Decay & $\tau$$^{(a)}$\\
    (keV)    &E$_\gamma$(keV) & (ps)&    (keV)    &E$_\gamma$(keV) & (ps)\\
\hline
2667& 563.8&$\le$ 1.4 &2975&1620&$\le$ 1.4\\
2837&1005&$\le$ 1.4 &3011&406.6&$\le$ 1.4\\
2860&757.2&$\le$ 1.4 &3024&1669&$\le$ 1.4\\
2864&490.2& $\le$ 0.55 &3203&1371&$\le$ 1.4\\
2867&397.7&$\le$ 1.4 &3210&740.2&$\le$ 1.4\\
2881&1049&$\le$ 0.55 &&&\\
\hline
\end{tabular}
\end{center}
\end{table}
As discussed previously (see sect.~\ref{level scheme}), 
most of these transitions have an E1 multipolarity. 
All the other transitions having an energy lower than 1~MeV are also assigned as
E1, since their transition strengths, calculated for
M1 or E2 multipolarities, would lead to unexpectedly large values.

\begin{figure}[h!]
\begin{center}
\includegraphics*[width=7cm]{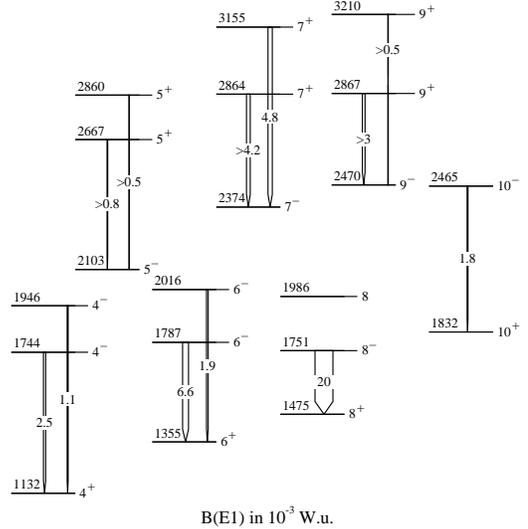}
\caption{States of $^{212}$Po decaying by enhanced E1 transitions, the width of each
arrow is proportional to the B(E1) value (given in 10$^{-3}$ W.u.).  
\label{levels_BE1}
}
\end{center}
\end{figure}

The values of the B(E1) reduced transition probabilities have 
been calculated, taking into account the experimental values 
of the $\alpha_{tot}$ coefficients (see table~\ref{ICC}). The results, 
given in table~\ref{E1transitions}, are displayed, as a function of
their location in the level scheme, in fig.~\ref{levels_BE1} and will
be discussed in Sect.~\ref{discuss_cluster}.
\begin{table}[!h]
\begin{center}
\caption{Values of the B(E1) reduced transition probabilities of the 
enhanced E1 $\gamma$-rays measured in this work.
}
\label{E1transitions}
\begin{tabular}{|c|c|c|}
\hline
 E$_\gamma$(E1)  & B(E1)$^{(a)}$  & B(E1)$^{(a)}$ \\
 (keV)	  	&($e^2fm^2$) 	 &(W.u.)\\
\hline
276.1		& 4.6(23) 10$^{-2}$	&2.0(10) 10$^{-2}$\\
398.0		&$\ge$ 7.0 10$^{-3}$       &$\ge$ 3.0 10$^{-3}$\\
406.6		&$\ge$ 6.5 10$^{-3}$       &$\ge$ 2.8 10$^{-3}$\\
432.3		& 1.5(3) 10$^{-2}$	&6.6(12) 10$^{-3}$\\
490.2		&$\ge$ 9.6 10$^{-3}$	&$\ge$ 4.2 10$^{-3}$\\
563.8		&$\ge$ 1.7 10$^{-3}$	&$\ge$ 7.6 10$^{-4}$\\
612.3		& 5.7(18) 10$^{-3}$	&2.5(7) 10$^{-3}$\\
633.3		& 4.1(11) 10$^{-3}$	&1.8(5) 10$^{-3}$\\
661.3		& 4.4(14) 10$^{-3}$	&1.9(6) 10$^{-3}$\\
740.2		&$\ge$ 1.1 10$^{-3}$	&$\ge$ 4.8 10$^{-4}$\\
757.2		&$\ge$ 1.0 10$^{-3}$	&$\ge$ 4.5 10$^{-4}$\\
780.4		& 1.1(6) 10$^{-2}$	&4.8(26) 10$^{-3}$\\
813.6		& 2.5(8) 10$^{-3}$	&1.1(3) 10$^{-3}$\\
\hline
\end{tabular}
\end{center}
$^{(a)}$ the number in parenthesis is the error in the last digit.
\end{table}

\subsection{Peculiarity of the $^{ 212}$Po level scheme obtained in this work}

A close examination of fig.~\ref{schema_popotwo} reveals that the same basic
structure arises three times, which expands above the I$^+$ yrast states with  
I~=~4, 6, 8 (the fact that this was not observed above the 2$^+_1$ state in
our experiment can be due to the trigger level used to record the events). 
This structure comprises:
\begin{itemize}
\item two I$^-$ states (drawn in magenta) which only decay to the I$^+$ yrast
state {\it via} an enhanced E1 transition. 
\item one (I~+~1)$^-$ (drawn in blue) which decays to the I$^+$ and (I~+~2)$^+$
yrast states, as well as to the two I$^-$  states (drawn in magenta). Moreover, in
two cases, the decay to the low-lying (I~+~2)$^-$ state is also observed.
\item two (I~+~1)$^+$ states (drawn in violet) which only decay to the 
(I~+~1)$^-$ states (drawn in blue) {\it via} an enhanced E1 transition. 
\end{itemize}
Such a feature gives confidence in the spin and parity assignments we have 
done using all the experimental results reported in the previous 
sections.

\subsection{$\alpha$ branching ratios}\label{alpha}

Several yrast states of $^{212}$Po were reported to partially decay 
{\it via} $\alpha$ emission to the ground state of $^{208}$Pb, as the 
corresponding $\alpha$ lines were clearly identified~\cite{ba78,li78,le80,es84}. 
The branching ratios given in column 4 of table~\ref{alpha_branching} 
are those quoted in the last evaluation 
of the nuclear spectroscopic information for the
$A$=212 nuclides~\cite{NDS}. It is noteworthy that the numbers for the 4$^+$, 
6$^+$, and 8$^+$ states are very large, they come from the analysis of the 
$\beta$ decay of the 9$^-$ isomeric state of $^{212}$Bi which populates excited 
states of $^{212}$Po (see ref.~\cite{es84}).
The value, $b_{\alpha}$($8^+$)$\sim 42 \%$, is well greater than the one
previously determined using $\alpha$-$\gamma$ coincidence~\cite{li78}, 
$b_{\alpha}$($8^+$)$\sim 6 \%$.  

\begin{table}[!h]
\caption[h]{\label{alpha_branching} 
$\alpha$ branching ratios ($b_{\alpha}$) of the first yrast states of $^{212}$Po
given in the last evaluation of the nuclear spectroscopic information for the
$A$=212 nuclides~\cite{NDS}. The new
values are evaluated from refs.~\cite{ba78,li78,le80} as well as our results 
(see text). The new values of T$^{\alpha}_{1/2}$ are given in the last column.}
\begin{center}
\begin{tabular}{|c|c|c|c|c|c|}
\hline
State  & I$^\pi$& T$_{1/2}$	& \multicolumn{2}{|c|}{$b_{\alpha}$ (\%)}  &T$^{\alpha}_{1/2}$ \\
(keV)       &     &    ns      &  from \cite{NDS}  &    new value   & ns\\
\hline
  0         &  $0^+$   &299(2) &  100          &  100      	     &299(2)\\
  727     &  $2^+$     &	&  0.033       &    0.033  	     &\\
  1132     &  $4^+$    &	&  $\sim 27$     & $\sim 0.5$ 		&\\
  1355     &  $6^+$    &0.76(21) &  $\sim 71$     &  $\sim 3(1)$  	&25(10)\\
  1475     &  $8^+$    &14.7(3)&  $\sim 42$     &   $\sim 3(1)$  	&490(150)\\
\hline
\end{tabular}
\end{center}
\end{table}

Large values of $\alpha$ branching ratios could easily be confirmed by 
intensity balances in our $\gamma$-spectra gated by at least one transition 
lying in the top of the level scheme. Indeed any $\alpha$ emission would lead 
to a loss of the total intensity ({\it i.e.} including electron conversion) of 
the transitions located below the $\alpha$ emitting state. From the analysis 
of such double-gated spectra, we did not confirm any large value of 
$b_{\alpha}$: the five yrast transitions (357~keV, 120~keV, 
223~keV, 405~keV and 727~keV) exhibit total intensities which are  constant 
within the error bars.

In order to go further, we have carefully examined the four papers published on
this subject~\cite{ba78,li78,le80,es84}. 
\begin{figure}
\begin{center}
\resizebox{0.2\textwidth}{!}{\includegraphics*{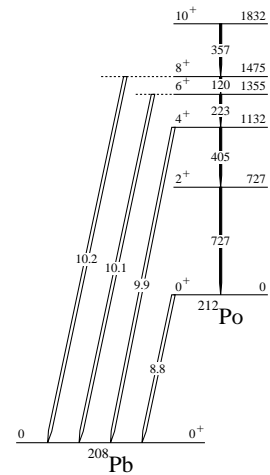}}
\caption{Schematic representation of the $\alpha$ decay-lines of 
$^{212}$Po~\cite{NDS}. Values of $\alpha$ energies are given 
in MeV, whereas those of $\gamma$-transition and level energies 
of $^{212}$Po are given in keV.
}
\label{alpha_fig}       
\end{center}
\end{figure}
Baidsen {\it et al.} \cite{ba78} measured three $\alpha$-lines of energies
$E_\alpha = $~9.89~MeV, 10.10~MeV, and 10.22~MeV (see fig.~\ref{alpha_fig}), 
having {\it relative} intensities 
of 8\%, 46\%, and 46\%, respectively. These $\alpha$ lines were interpreted as 
the $\alpha$-decays of three excited states of $^{212}$Po, lying above 
1~MeV
excitation energy. Simultaneously, from results of $\alpha - \gamma$ 
coincidences and half-life measurements, Lieder {\it et al.}~\cite{li78} 
established (i) a 10.18~MeV $\alpha$ line deexciting a state with 
T$_{1/2}$=14.2(24)~ns in $^{212}$Po and (ii) several $\gamma$ transitions correlated 
in time 
with the 8.78~MeV $\alpha$ line emitted by the ground state of $^{212}$Po. 
This led to the identification of the $4^+$, $6^+$ yrast states (from the
cascade 223-405-727) and the 14.2~ns isomeric state was interpreted as the 
$8^+$ yrast state, at 1423 $\pm$ 30 keV from the energy balance of the $\alpha$
and $\gamma$ lines. Moreover, the intensities of the $\alpha$ and $\gamma$
transitions deexciting this new $8^+$ state were found to be $I_\alpha$=0.06(1)
and $I_\gamma$=0.94(1), respectively.

This has been confirmed two years later
from $\alpha - \gamma$ coincidences~\cite{le80}, particularly the 120~keV
transition deexciting the $8^+$ yrast state was measured for the first time and
the $\alpha$ transition deexciting the $8^+$ yrast state was clearly resolved
from the one deexciting the $6^+$ yrast state, both with similar intensity, 
whereas the line associated with the $4^+$ state was well weaker 
(in agreement with results of ref.~\cite{ba78}, above-mentioned). 

At this stage, we can conclude that the 10.18~MeV $\alpha$ line of 
ref.~\cite{li78} is the doublet (see fig.~\ref{alpha_fig}) of the two lines 
at 10.1~MeV and 10.2~MeV given in refs.~\cite{ba78,le80}. Thus
the $I_\alpha$ value attributed at that time to the $8^+$ yrast 
state (6\%) has to be divided
by 2, the second part being ascribed to the $\alpha$ decay of the $6^+$ yrast 
state. These results, which are quoted in column 5 of 
table~\ref{alpha_branching}, are now 
in quantitative agreement with our intensity balances. 

The result of the $\beta$ decay of the 9$^-$ isomeric state of 
$^{212}$Bi made by Eskola {\it et al.}~\cite{es84}, which was adopted in the last
compilation~\cite{NDS}, indicates very strong components of 
$\alpha$ emission from the yrast states of $^{212}$Po which were neither found 
in the previous experiment~\cite{li78} nor in our data set. 
Moreover, the  $\alpha$ spectrum displayed in that paper does
not corroborate such a result, as the intensity of the $\alpha$ group of 
$\sim 10$~MeV is less than the one corresponding to the $^{212}$Po ground 
state decay, by more than one order of magnitude. So we can wonder whether 
there is a misprint in ref.~\cite{es84}, the number of $\alpha$ decay of 
the excited states being reversed with the one of the ground state.  

The new values of the $\alpha$ partial half-lives of the yrast states are 
given in the last column of 
table~\ref{alpha_branching}, they will be discussed in the next section in
connection with the cluster structure of the states.

\section{Discussion}\label{discussion}

It would be very straigthforward to predict the structure of $^{212}$Po,  
with four nucleons outside $^{208}$Pb, in terms of shell-model (SM) configurations, 
using the single-particle states in the mean field of the doubly-magic core.   
Nevertheless as the four nucleons are two protons and two neutrons, we have also
to take into account the four-particle correlation leading to the formation of the
$\alpha$-particle.  
In this section, we aim to isolate the effects of four-particle correlations 
on the properties of the excited states of $^{212}$Po (excitation energy, spin and
parity, deexcitation modes). For that purpose, we use
{\it empirical} arguments from the comparison with the neighbouring isotopes and
isotones which are mainly described within SM configurations.   

\begin{figure}[h!]
\begin{center}
\includegraphics*[width=7.5cm]{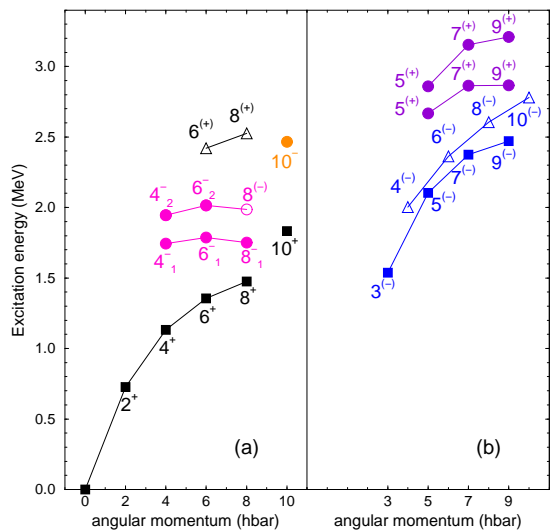}
\caption{(color online) Excitation energy as a function of angular
momentum of most of the $^{212}$Po states observed in the present
work (they are displayed in figs.~\ref{schema_popo1} and 
\ref{schema_popotwo} using the same colors). 
Each state drawn with a filled circle (magenta, orange, and violet) decays 
by an enhanced E1 transition towards the state with the same angular momentum.  
\label{States_212Po}
}
\end{center}
\end{figure}

Figure~\ref{States_212Po} displays most of the $^{212}$Po states observed 
in the present work, grouped as a function of their underlying structures 
which are reviewed by turns:
\begin{itemize} 
\item the yrast positive-parity states (in black) are discussed 
in terms of two-particle excitation, as their excitation energy as a function of
angular momentum follows the typical curve expected in case of pair
breaking~\cite{ri80}. Nevertheless several features point out the 
simultaneous effect of four-particle correlations (sects.~\ref{yrast_SM_Cluster} 
and \ref{discuss_alpha})
\item the negative-parity states (in blue) involve the coupling of the low-lying
3$^-$ octupole vibration to the excitation of the valence neutrons 
(sect.~\ref{discuss_octupol})
\item the non-natural parity states (in magenta, orange, and violet) are the 
fingerprints
of the '$\alpha$~+~$^{208}$Pb' structure (sect.~\ref{discuss_cluster}).
\end{itemize}

\subsection{Is the nucleon-pair breaking the unique excitation mode of
$^{212}$Po at low energy?\label{yrast_SM_Cluster}}
\begin{figure}[h!]
\begin{center}
\includegraphics*[height=6.0cm]{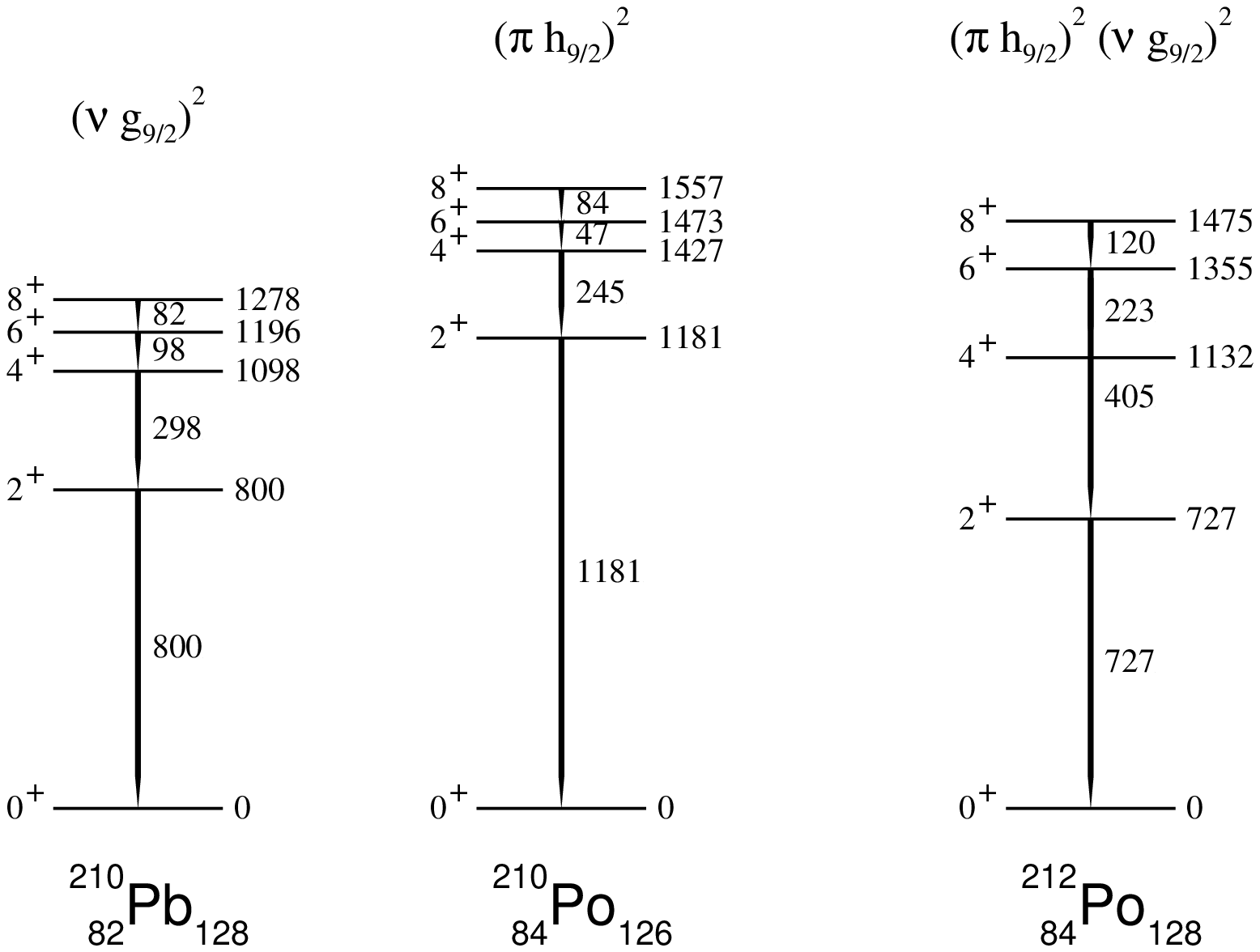}
\includegraphics*[height=6.0cm]{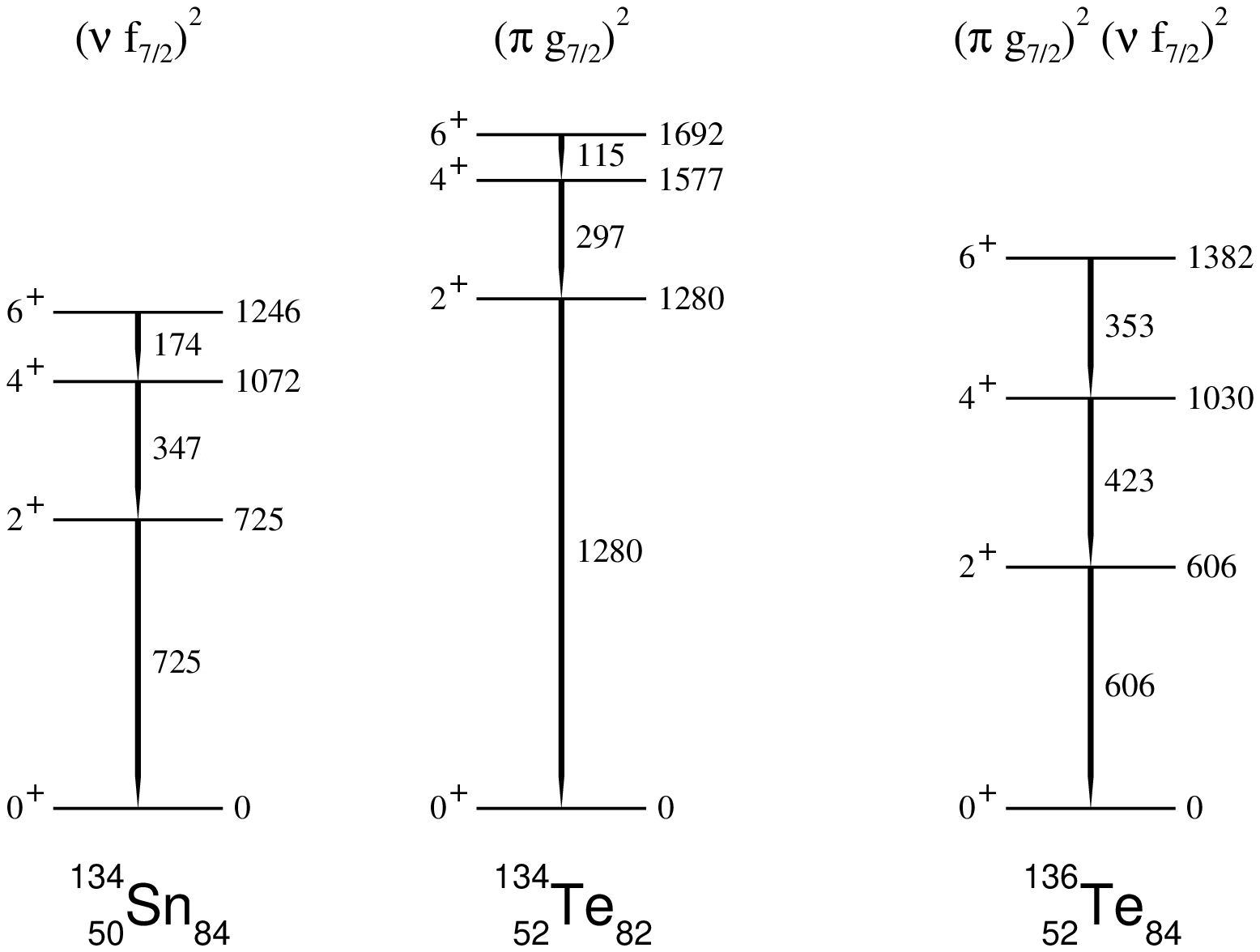}
\caption{{\bf Top}: First excited states of $^{210}$Pb, $^{210}$Po,  
and $^{212}$Po, the three 8$^+$ levels are isomeric
(T$_{1/2}$ = 201(17)~ns, 98.9(25)~ns, and 17.1(2)~ns, 
respectively).
{\bf Bottom}: First excited states of $^{134}$Sn, $^{134}$Te,  
and $^{136}$Te. The  6$^+$ levels of $^{134}$Sn and $^{134}$Te 
are isomeric (T$_{1/2}$ = 80(15)~ns and 164.1(9)~ns, 
respectively).
\label{PbPoSnTe}}
\end{center}
\end{figure}

The first excited states of $^{210}_{}$Pb$^{}_{128}$ and 
$^{210}_{~84}$Po are text-book examples of residual interaction 
in a two-particle configuration, namely $(\nu g_{9/2})^2$ and 
$(\pi h_{9/2})^2$, respectively.
While the energy interval of the whole multiplet extends over 
1557 keV for the two-proton case, the one of the two-neutron
configuration is more  compressed (1278 keV) (see 
the top part of fig.~\ref{PbPoSnTe}). That is directly
connected to the energies of the residual interactions, as discussed
below. 

The structure of the first levels of $^{212}_{~84}$Po$_{128}$  
(see fig.~\ref{PbPoSnTe}) would be described as a
superposition of the excitations known in its two even-even 
neighbours. The low energy of the 2$^+_1$ state (727 keV) indicates 
that the $(\nu g_{9/2})^2$ part of its  wave function is likely 
stronger than the $(\pi h_{9/2})^2$ one. The fact 
that the neutron pair breaks more easily than the proton one beyond 
the doubly-magic nucleus $^{208}$Pb, is corroborated by the empirical 
values of the proton (neutron) pairing energies extracted 
from the measured binding energies of the ground states of the 
odd-Z (odd-N) nuclei surrounding $^{212}$Po. In order to avoid the
effect of the extra binding energy of the doubly-magic nucleus 
$^{208}$Pb, we use the three-point formula~\cite{sa98}, defining a local
average of the masses of two odd-A nuclei which is compared with
the observed mass of the even-even $^{212}$Po nucleus:
\begin{equation}
\Delta_p^{(3)}(84) = BE(84,128)-{\frac{1}{2}}~ 
[BE(83,128)+BE(85,128)] 
\end{equation}
\begin{equation}
\Delta_n^{(3)}(128) = BE(84,128)-{\frac{1}{2}}~
[BE(84,127)+BE(84,129)] 
\end{equation}
The neutron pairing energy, 
$\Delta_n^{(3)}$= 826(2)~keV, is lower than the proton one,
$\Delta_p^{(3)}$= 1150(3)~keV,
by $\sim$ 320 keV\footnote{The same difference is obtained when computing
$\Delta_p^{(3)}(84)$  associated to $^{210}$Po and
 $\Delta_n^{(3)}(128)$ associated to $^{210}$Pb.}. 

Noteworthy is the fact that a similar situation occurs in other
nuclei located just above the doubly-magic core of $^{132}$Sn
(see the bottom part of fig.~\ref{PbPoSnTe}). Firstly, the two-neutron 
configuration of $^{134}$Sn leads to a more compressed level scheme 
than the two-proton configuration of $^{134}$Te. 
Secondly, the low energy of the
2$^+_1$ state of $^{136}$Te occurs in concert with a low value of
the neutron pairing energy. The three-point formula give
$\Delta_p^{(3)}(52) =$ 1320(60)~keV and 
$\Delta_n^{(3)}(84) =$ 730(82)~keV. Thus the 
neutron pair breaks more easily than the proton one and 
the 2$^+_1$  wave function of $^{136}$Te should contain a large
component of neutron excitation. Such a scenario is strengthened 
by the B(E2; 0$^+$ $\rightarrow$ 2$^+_1$) values, which have been 
recently measured by Coulomb excitation in inverse kinematics 
\cite{Radf05}. The value obtained in $^{134}$Te 
(4.7(6)~W.~u.) is stronger than the $^{134}$Sn one 
(1.4(2)~W.~u.), as expected from their different excitation
processes (breaking of a proton pair versus a neutron pair).
Moreover the result obtained in $^{136}$Te  
has confirmed that the wave function of its 2$^+_1$ state mainly comes 
from neutron excitation~\cite{te02,co07}.

\begin{figure}[h!]
\begin{center}
\includegraphics*[height=6.0cm]{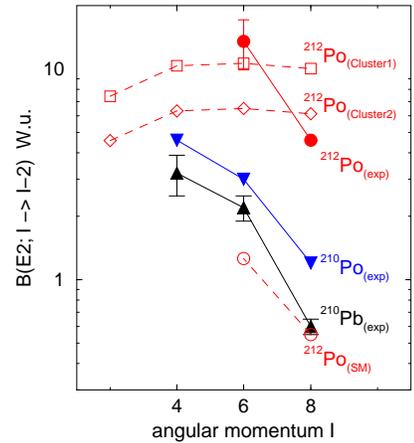}
\caption{(color online) Experimental values of the reduced transition probabilities 
B(E2; I $\rightarrow$ I-2)
of the first yrast states of $^{210}_{~82}$Pb$^{}_{128}$ (ref.~\cite{NNDC}), 
$^{210}_{~84}$Po$^{}_{126}$ (ref.~\cite{NNDC}), and 
$^{212}_{~84}$Po$^{}_{128}$ (ref.~\cite{po87} and this work). The theoretical 
values of $^{212}$Po come from shell-model approach~\cite{po87}, 
and from $\alpha$-$^{208}$Pb cluster models (Cluster1 from
ref.~\cite{ho94}, Cluster2 from ref.~\cite{bu96}). 
\label{BE2_PoPb}}
\end{center}
\end{figure}

Unfortunately the leading role of neutron excitation in the wave 
function of the 2$^+_1$ state of $^{212}$Po cannot be validated by
the value of the B(E2) reduced transition, as it is not
known. Nevertheless the comparison of the B(E2) values within
the seniority-2 states of the three neighbours is especially 
instructive in this respect (see fig.~\ref{BE2_PoPb}).   
As expected, the values of $^{210}$Po are larger than those of 
$^{210}$Pb, owing to  their different excitation processes 
(proton pair {\it versus} neutron pair). Similarly 
the values of the 6$^+_1$ state and 8$^+_1$ state of $^{212}$Po, 
calculated using the shell-model approach~\cite{po87} 
(see the empty circles in fig.~\ref{BE2_PoPb}) are low since these
states are predicted to come mainly from neutron excitation, 
as above-mentioned. 
On the other hand, the measured values 
(ref.~\cite{NNDC} and this work for the new 
values of the $\alpha$~/~$\gamma$ branching ratios) are 
greater than the calculated ones by one order of magnitude, implying 
that these 6$^+_1$ and 8$^+_1$ states are complex. Even though their 
energies are rather well reproduced from the sole two-neutron 
excitation, other excitation modes have to be taken into account, 
such as the $\alpha$ + $^{208}$Pb cluster states which lead to very 
large values of B(E2) as displayed in fig.~\ref{BE2_PoPb}. 

Before closing this section, we want to recall what SM 
configurations were attributed to the high-spin yrast 
states~\cite{po87}. The excitation of one neutron to 
the second orbit above the magic number gives 
the $\nu g_{9/2} \nu i_{11/2}$ configuration, whereas the excitation of
one nucleon to the first intruder orbit above the magic numbers gives 
either $\nu i_{11/2} \nu j_{15/2}$ or $\pi h_{9/2} \pi i_{13/2}$. 
Thus, we expect three multiplets of
states with spin values I$^\pi$~=~1$^+$ to 10$^+$, ~2$^-$ to 13$^-$, and
2$^-$ to 11$^-$ respectively. Since the two nucleons have the same
character (particle-particle), the energies of the states having the 
extreme spin values (1$^+$ and 10$^+$, ~2$^-$ and 13$^-$, and
2$^-$ and 11$^-$ respectively) are the most lowered by the residual
interactions. That provides reliable assignments for the yrast states 
located at 1832~keV (10$^+$), 2769~keV (13$^-$), and 2409~keV (11$^-$). 
It is worth pointing out that in the framework of the SM approach, 
the levels with I$^\pi$~= ~4$^-$ to 9$^-$ are expected above
2409~keV, the energy of the lowest negative-parity level. 
As for the higher spin values (I $>$ 10$^+$), they are obtained from 
the simultaneous excitation of the two pairs of nucleons, 
$(\nu g_{9/2})^2 \otimes (\pi h_{9/2})^2$ (I$_{max}$ = 16$^+$) or
$(\nu g_{9/2} \nu i_{11/2}) \otimes (\pi h_{9/2})^2$ (I$_{max}$ = 18$^+$).
\subsection{Are the $\alpha$-decays of $^{212}$Po enhanced ?\label{discuss_alpha}}
Besides its ground state which decays by $\alpha$ emission, $^{212}$Po
has several excited states which are also $\alpha$ emitters, some of 
the partial half-lives are given in the last column of 
table~\ref{alpha_branching} (sect.~\ref{alpha}). The present section is 
devoted to the comparison
of the $\alpha$-emission probabilities of $^{212}$Po to those of the
neighbouring isotopes and isotones in order to show {\it
empirically} that the cluster structure, 
$\alpha$ + $^{208}$Pb, plays an important role in the process.

The well-known Geiger-Nuttall plots show the correlation 
between the decay half-life and the energy of the emitted $\alpha$:
\begin{equation}\label{Geiger-Nuttall}
log_{10}T_{1/2}=a E^{-1/2}_\alpha +b
\end{equation} 
The parameters are usually obtained by fitting the data 
associated to the ground states of nuclei of each isotopic series.
It is important to note that when an isotopic series 
crosses a neutron 
magic number, such as $N$=126, the results separate into
two groups, each one having its own parameters~\cite{bu90}. 
In parallel, intensive works have been done in order to obtain a 
{\it universal} description of $\alpha$ decays, as well as cluster 
radioactivity (see, for instance, the new law recently 
published in ref.~\cite{qi09}). 

\begin{figure}[h!]
\begin{center}
\includegraphics*[width=7cm]{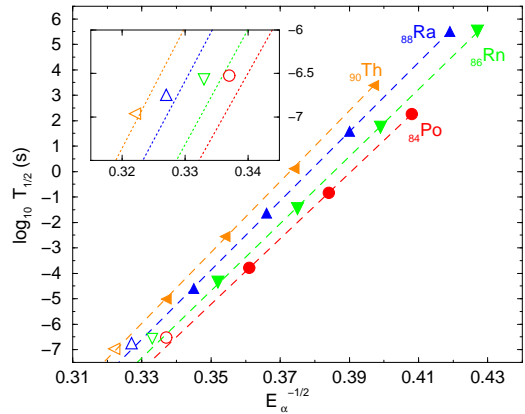}
\caption{(color online) 
Geiger-Nuttall relations for four sets of isotopes ($_{84}$Po, 
$_{86}$Rn, $_{88}$Ra, $_{90}$Th) having $N$=130-136 (filled symbols). 
The experimental data for$N$=128 are drawn with empty symbols. 
The inset gives a zoom on the four fits to emphasize the behaviour 
of the $N$=128 isotones, $^{218}$Th (triangle left), $^{216}$Ra 
(triangle up), $^{214}$Rn (triangle down) and $^{212}$Po (circle).
\label{alphas_PoRnRaTh}}
\end{center}
\end{figure}

For our purpose, we have only used the data associated to 
nuclei with $N$=130-136 of four isotopic series ($_{84}$Po, 
$_{86}$Rn, $_{88}$Ra, $_{90}$Th), excluding the nuclei of interest, 
the $N$=128 isotones. The values of the two  coefficients of the 
Geiger-Nuttall relations (eq.~\ref{Geiger-Nuttall}) are
given in table~\ref{coeff_Geiger-Nuttall} and the fits are drawn in 
fig.~\ref{alphas_PoRnRaTh}.
\begin{table}[!h]
\caption[h]{\label{coeff_Geiger-Nuttall}  
Values of the two coefficients of the Geiger-Nuttall
relations (eq.~\ref{Geiger-Nuttall}) describing the $\alpha$-decay 
half-lives of the ground states of nuclei having N $\ge$ 130. The last
column gives the value of the hindrance factor (HF) of each 
$N$=128 isotone (see text). 
}
\begin{center}
\begin{tabular}{|cc|cc|c|}
\hline
	& N & a &  b & HF(N=128) \\
\hline
$_{90}$Th & 130-136   & 139.4$\pm$1.4   & -51.98$\pm$0.50  & 1.4\\
$_{88}$Ra & 130-136   & 136.1$\pm$0.9   & -51.50$\pm$0.35  & 1.8\\
$_{86}$Rn & 130-136   & 131.9$\pm$1.5   & -50.84$\pm$0.58  & 2.2\\
$_{84}$Po & 130-134   & 128.8$\pm$0.4   & -50.30$\pm$0.16  & 2.3\\
\hline
\end{tabular}
\end{center}
\end{table}
The half-life values extend over 10 orders 
of magnitude, thus we  give a zoom on the
behaviour of the four $N$=128 isotones, in the inset of
fig.~\ref{alphas_PoRnRaTh}. 
Their measured $\alpha$ half-lives 
are greater than the values expected from the fits 
(each empty symbol lies above the
corresponding line), allowing us to calculate
the hindrance factors (HF), i.e. the ratio of the experimental half-life 
and the extrapolated value (given in the last column of 
table~\ref{coeff_Geiger-Nuttall}). 
Equation~(\ref{Geiger-Nuttall}) mainly accounts for the $\alpha$ 
tunnelling through the potential barrier, but does not take into 
account any variation of the probability of $\alpha$ formation 
in the parent nuclei. Actually the quality of the fits shown in 
fig.~\ref{alphas_PoRnRaTh} proves that this probability remains almost
the same along an isotopic series as soon as {\it at least} two pairs of 
neutrons are available (i.e. N$\ge$ 130). On the other hand, the 
hindrance factors of the N~=~128 isotones (HF $>$ 1) mean that 
it is more difficult to form an $\alpha$ 
particle in the parent nuclei which have {\it only} two neutrons above 
a magic number. As seen in the last column of 
table~\ref{coeff_Geiger-Nuttall}, the HF values increase 
regularly as $Z$ is decreasing from $Z$~=~90 to 86, while the value
for $^{212}_{~84}$Po is almost the same as the $^{214}_{~86}$Rn one.  
This means that the $^{212}$Po half-life is smaller than expected 
from the behaviour of its heavier isotones. Such a result is likely
due to a greater value  of the probability of $\alpha$ 
formation in the $^{212}$Po ground state, a sign of its cluster 
structure, $\alpha$ + $^{208}$Pb. It is worth recalling that
the amount of clustering predicted by an hybrid model comprising 
both shell and cluster configurations~\cite{va92} is high (30\%). 

Using the partial half-lives given the last column of 
table~\ref{alpha_branching}, the HF values of two excited 
states of $^{212}$Po can be computed as well,
HF$(6^+_1)=1.8 \times 10^2$  and HF$(8 ^+_1 )=6.0 \times 10^3$. 
In those cases,
because of the conservation of the total angular momentum, the alpha
emission occurs at non-zero value of orbital angular momentum, giving
rise to a centrifugal potential which increases the potential barrier
and slows down the tunnelling process~\cite{pr75}. 
It would have been interesting to make quantitative comparisons with 
transitions involving similar variations of angular 
momentum in neighbouring nuclei, in order to look for hint of cluster 
structure in these excited states. Unfortunately 
the number of cases available in the literature~\cite{NNDC} is too 
scarce. 

Finally we have to note that the new values of the $\alpha$ branching
ratios of the 4$^+_1$, 6$^+_1$ and 8$^+_1$ states given in 
table~\ref{alpha_branching} are in good agreement with the predictions of
the $\alpha$-cluster models of $^{212}$Po (see for instance 
refs.~\cite{ho94,bu96}).

\subsection{Excitations due to an octupole mode in
$^{212}$Po}\label{discuss_octupol}

The octupole mode is well known in the heavy nuclei. 
It gives rise to the lowest excited state of $^{208}$Pb, with
E~=~2.614 MeV and I$^\pi$~=~3$^-$, the E3 transition towards the
ground state being collective, B(E3)~=~33.8(6)~$W.u.$. The structure 
of this octupole vibration comprises many coherent particle-hole 
excitations across the two magic gaps at $Z=82$ and $N=126$ involving
several single-particle orbits with $\Delta l=3$. When the Fermi levels
are located well above these two gaps, namely for $N \sim$ 134 and 
$Z \sim$ 88, the octupole mode is enhanced because of
strong octupole couplings between two pairs of orbits 
($\nu g_{9/2}- \nu j_{15/2}$ in the one hand, 
$\pi f_{7/2}- \pi i_{13/2}$ in the other hand).

In between, there are various mixings of the 3$^-$ excitation of the 
$^{208}$Pb core and  the single-particle excitations from these pairs 
of orbits. For instance, in $^{210}$Pb, the 3$^-$ collective 
excitation of the core and the 3$^-$ state of the 
$\nu g_{9/2}- \nu j_{15/2}$ configuration, expected close in energy, 
are so strongly mixed that the measured octupole strength is split 
into two states 1~MeV apart (1869~keV and 2828~keV). They account 
for the full B(E3) intensity which has been observed in $^{208}$Pb 
and as they share this intensity in the ratio 2:1~\cite{el71}, the energy 
of the 3$^-$ collective excitation can be computed, E~$\sim$~2.2~MeV. 

\begin{figure}[h!]
\begin{center}
\includegraphics*[width=6cm]{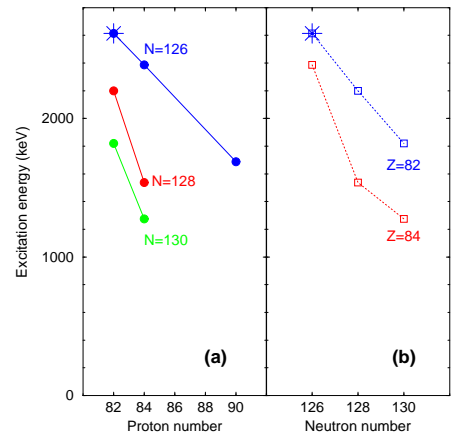}
\caption{(color online) Evolution of the excitation energy of the 3$^-$ collective 
excitation when adding nucleon pairs 
to the $^{208}$Pb doubly-magic core (marked with a star), as a function
of the proton number (a) and of the neutron number (b). 
\label{E_octupole}}
\end{center}
\end{figure}

Moreover, one can expect that adding nucleon pairs to $^{208}$Pb
would result in a greater softness of the collective 3$^-$ octupole mode, 
with a lowering of its energy well below 2.614 MeV. 
This was put forward in order to explain the behaviour of the E3 strength
in the $N=127$ isotones, in the framework of particle-octupole vibration
coupling~\cite{dr89}.
The evolution of the excitation energy of the 3$^-$ collective state 
is displayed in fig.~\ref{E_octupole}, both as a function of $Z$ and as 
a function of $N$. 
The decrease in energy is linear when adding proton pairs or neutron 
pairs to the doubly-magic core. The slope is $\sim$ 220 keV per 
proton pair (see the blue solid line in fig.~\ref{E_octupole}a)
and $\sim$ 400 keV per neutron pair (see the blue dotted line in 
fig.~\ref{E_octupole}b). On the other hand, the decrease is 
stronger when adding both proton pairs and neutron pairs (see the slopes
of the red or green lines). 

The case of $^{212}$Po
seems peculiar as shown by the break in the red dotted line (see 
fig.~\ref{E_octupole}b). When assuming a linear behaviour as a 
function of neutron number for $Z=84$, the energy of its 3$^-$ state 
would be 300~keV higher than measured. This is more likely the result 
of some mixing. It could involve the $\nu g_{9/2}- \nu j_{15/2}$
configuration as in $^{210}$Pb, discussed above. Nevertheless the 
effect of the 
cluster structure, $\alpha$ + $^{208}$Pb, should not be neglected.
It is worth pointing out that the 10.302~MeV $\alpha$-ray,
observed in ref.~\cite{le80}, that defines an excited state 
at 1547(10)~keV~\cite{NNDC} could be assigned 
to the 3$^-$ state at 1537~keV. The fact that the $\alpha$-emission can
compete with the $\gamma$ decay would be a sign of a cluster part in 
the wave function of the 3$^-$ state. 

At higher excitation energy, we can expect negative-parity states 
from the coupling of the 3$^-$ collective state 
to the valence nucleon excitation. In principle, several multiplets 
of states should be grouped at excitation energies given by the 3$^-$ 
energy and the positive-parity yrast spectrum. Because of the residual 
interactions, the multiplets are no longer degenerated in energy and the
highest-spin states of each multiplet would be the lowest ones. Such a behaviour is
well known in the region of $^{146}$Gd, where the octupole 
excitation also plays an important role.
   
\begin{figure}[h!]
\begin{center}
\includegraphics*[width=8cm]{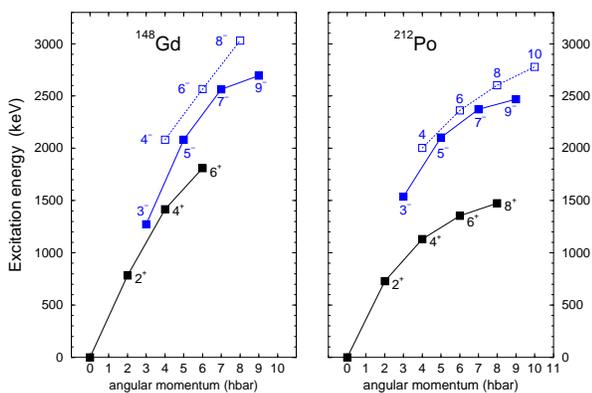}
\caption{(color online) Left: Excitation energy, as a function of angular
momentum, of some states of$^{148}_{~64}$Gd$^{}_{84}$, which can be
interpreted in terms of spherical shell-model configurations (black symbols) 
and couplings of two valence 
nucleons to one octupole phonon (blue symbols), from ref.~\cite{pi90}. 
Right: Similar states of $^{212}_{~84}$Po$^{}_{128}$ observed in 
our experiment. 
\label{ShellModelStates_212Po}
}
\end{center}
\end{figure}

The yrast states of  $^{148}$Gd with two valence neutrons
outside the $^{146}$Gd core nucleus have been identified and discussed
in terms of spherical shell model and couplings of two valence 
nucleons to one octupole phonon~\cite{pi90}. The four lowest yrast
states up to 6$^+$ arise from the $\nu f_{7/2}^2$ configuration (see
the left part of fig.~\ref{ShellModelStates_212Po}). The coupling of
this configuration to the $^{146}$Gd 3$^-$ core phonon gives rise to a
lot of negative parity levels. Some of them are mixed with the states   
from the excited $\nu f_{7/2}^1 \nu i_{13/2}^1$ configuration which
gives a
multiplet of eight states (3$^-$~$\le$~I$^\pi$~$\le$~10$^-$). As a result, the
first states with negative-parity display a characteristic behaviour,
one branch is formed by the first states having the odd values of angular
momentum, while the first states having the even values are located slightly
higher in energy (see the blue symbols in the left part of 
fig.~\ref{ShellModelStates_212Po}). Lifetimes of excited states in $^{148}$Gd
were measured ~\cite{po03b}: all the B(E1) values of the transitions linking the
negative-parity states with odd spin values to the positive-parity yrast 
states are around 10$^{-5}$-10$^{-6}$ W.u., i.e. within the "normal" range.  

It is worth pointing out that the two neutron shells involved in 
$^{148}$Gd have their counterparts in $^{212}$Po, with one unit larger
angular momenta $l$ and $j$ ( $\nu f_{7/2}$ $\rightarrow$ $\nu h_{9/2}$
and $\nu i_{13/2}$ $\rightarrow$ $\nu j_{15/2}$). This allows us to 
make reliable extrapolations to the case of $^{212}$Po. The 
comparison between the left and the right 
parts of fig.~\ref{ShellModelStates_212Po} indicates unambiguously that
the branch comprising the 5$^-$, 7$^-$, and 9$^-$  states comes mainly from 
the couplings of two valence nucleons to one octupole phonon. Moreover,
four states with even-spin values (at 2003, 2363, 2604, and 2780~keV), 
which decays the odd-$I$ states, 
probably form the second branch of negative-parity states in $^{212}$Po
as observed in $^{148}$Gd (see the blue open squares 
in fig.~\ref{ShellModelStates_212Po}).

\subsection{Cluster structure of the low-lying states of
$^{212}$Po}\label{discuss_cluster} 

Many theoretical works were devoted to the calculations of low-lying
states of $^{212}$Po assuming an {\it unmixed} $\alpha$-$^{208}$Pb cluster
structure (see for instance refs.~\cite{bu94,ho94,oh95,bu96}), the
potential being  either phenomenological with
different forms (such as
square well, cosh potential, or a modified Woods-Saxon function), or
obtained from a double folding procedure (i.e. the real
part of the optical potential giving a
good description of the $\alpha$~+~$^{208}$Pb scattering). In both
cases, the numerical values of some free parameters were determined to
reproduce the energies of the yrast states of $^{212}$Po, either
exactly~\cite{bu94,ho94,oh95} or at the best~\cite{bu96}.
It is worth pointing out that without these 
free parameters, all the predicted spectra of the yrast states 
are inverted (the highest-spin states lying below the lowest-spin 
ones, at variance with the experimental data). 
The main success of these calculations (as mentioned above, 
at the end of sect.~\ref{yrast_SM_Cluster} and fig.~\ref{BE2_PoPb}) 
is the better description of the B(E2) values inside the 
ground-state band, indicating that the wave-function of the yrast 
states having even-I and positive-parity values do possess some 
cluster content.

\subsubsection{Enhanced E1 transitions}
A major result of the present work is the observation of many 
strongly enhanced E1 transitions connecting several excited states 
to the yrast ones, having the same spin values (see fig.~\ref{levels_BE1}). 
Moreover as outlined above (at the end of sect.~\ref{yrast_SM_Cluster}), 
non-natural parity states, such as 4$^-$, 6$^-$, and 8$^-$, are not expected 
below 2.4~MeV excitation energy in the SM approach.

\begin{figure}[h!]
\begin{center}
\includegraphics*[width=7cm]{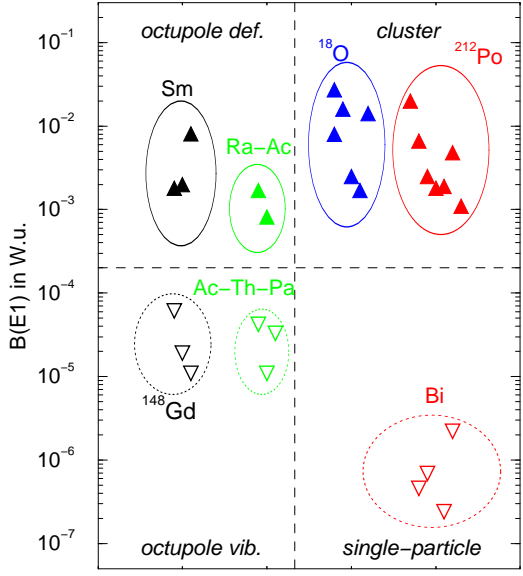}
\caption{(color online)\label{BE1_systematique} 
Values of B(E1) transition rates (in Weisskopf
units) for typical cases: Transitions in octupole-deformed  
and octupole-vibrational nuclei, transitions between 
single-particle states, and 
transitions involving cluster states.
}
\end{center}
\end{figure}

Enhanced B(E1) values are commonly found in nuclei exhibiting an 
{\it electric dipole moment}, such as 
light nuclei described in terms of a bimolecular system rotating
about its center of mass, or heavy nuclei displaying 
octupole deformation~\cite{ah93,buna96}. Some typical values  
are shown in fig.~\ref{BE1_systematique} using filled symbols 
(Sm isotopes with $N \sim 90$, $^{225}$Ra and $^{225}$Ac~\cite{NNDC}, 
cluster states of $^{18}$O~\cite{ga91}). They 
are one order of magnitude greater than those of transitions measured 
in octupole-vibrational nuclei, such as $^{148}$Gd, $^{225}$Ac, 
$^{231}$Th, $^{231}$Pa (shown with empty symbols) 
and three orders of magnitude greater than those of single-particle 
transitions in Bi isotopes.  

More generally, when a nucleus
clusterizes into fragments with different charge to mass ratios, 
the center of mass does not coincide any more with the center 
of charge, and a sizeable static E1 moment may arise in the intrinsic
frame~\cite{ia85}. The dipole moment is 
\begin{equation}\label{dipolemoment}
D = e(Z_1+Z_2)[\frac{Z_2}{(Z_1+Z_2)} - \frac{A_2}{(A_1+A_2)}] \times q
\end{equation}
where the parameter $q$ is the distance between the centers of mass of the two
fragments (see Eq.~(1) of Ref.~\cite{ia85}). 
Assuming two spheres in contact, the dipole moment of the
$\alpha$+$^{14}$C system 
is 0.9~$e$fm and very large values of E1 rates have been found in 
$^{18}$O~\cite{ga91} (see the blue symbols in fig.~\ref{BE1_systematique}). 
In the same way, the value of the 
$\alpha$+$^{208}$Pb system is as large as 3.7~$e$fm, which would 
lead to very enhanced E1 transitions. 
It is worth noting that the E1 rates measured in the present
work (shown with the red filled symbols in fig.~\ref{BE1_systematique})
are definitely located in the top part of the figure. This shows the  
'$\alpha$+$^{208}$Pb' structure of the even-I negative-parity states, 
which can be considered as {\it pure} cluster states since they are too 
low in energy to be accounted for using the SM approach (see the end 
of sect.~\ref{yrast_SM_Cluster}).   

One way of characterizing the cluster content of states, which  has 
been commonly done for the light nuclei, is measuring their
$\alpha$-decay width. Because of their non-natural$\alpha$ parity, the $\alpha$-decay
of the $^{212}$Po cluster states to the $^{208}$Pb ground state is expected 
to be forbidden (parity violation), although the energy factor would 
be very favourable. On the other hand, the $\alpha$-decay of the $^{212}$Po 
cluster states to the 3$^-_1$ excited state of $^{208}$Pb is allowed. 
Nevertheless the available energy for such an $\alpha$ emission is
so low that the $\alpha$-decay cannot compete with the very 
enhanced $\gamma$-decays, measured in the present work.  

\subsubsection{A possible model}

\begin{figure}[!h]
\begin{center}
\includegraphics[width=6cm]{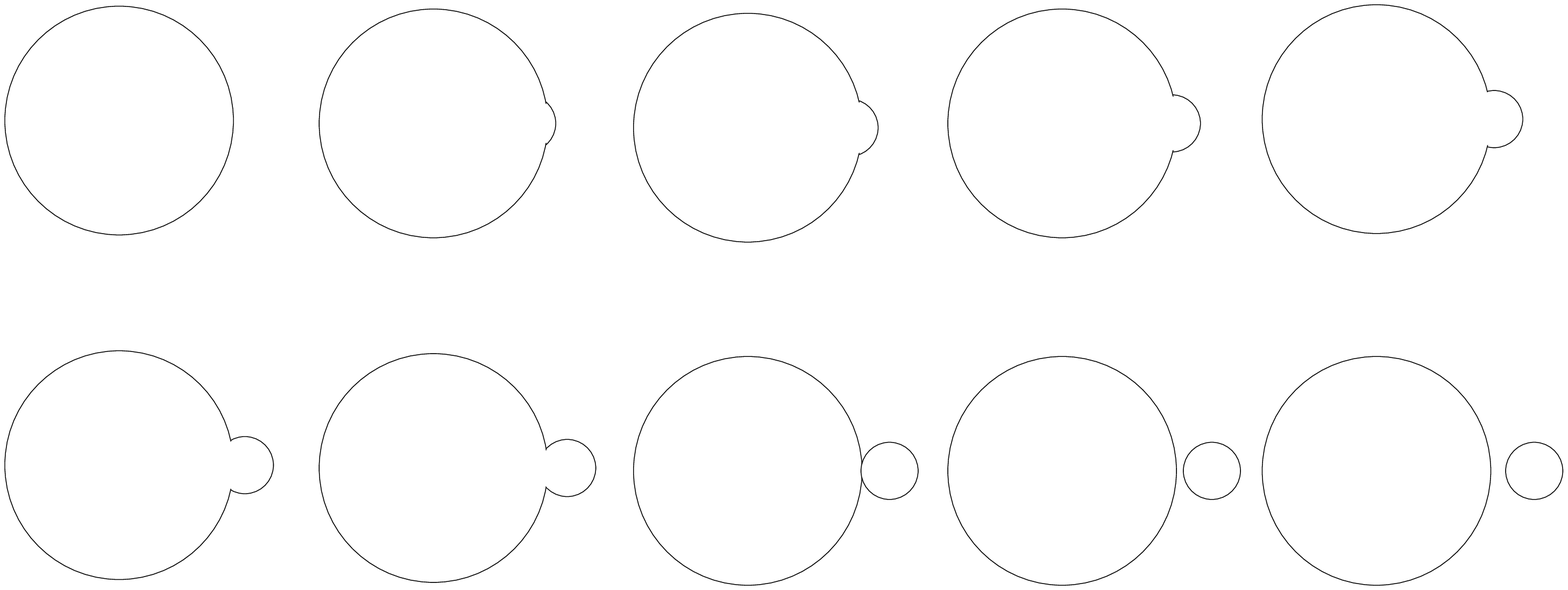}
\caption{Typical 
$^{208}$Pb $+ \alpha$ shapes as a function of the distance between their 
center of mass.
}
\label{pepin}       
\end{center}
\end{figure}

Following the ansatz of Ref.~\cite{va92} dealing with the ground state of
$^{212}$Po and taking into account the conclusions of the previous sub-sections, 
we suppose that the wave function of {\it every} excited state of $^{212}$Po can 
be written as the sum of two parts: 

\begin{equation}\label{fo}
\Psi^{tot}(I^{\pi}) = a~\Psi^{SM}(I^{\pi}) + b~\Psi^{cluster}(I^{\pi})
\end{equation}

\noindent
where $\Psi^{SM}$ stands for the pure shell-model piece and $\Psi^{cluster}$ 
takes account of the $\alpha$ clustering. This is the same as in our previous
work~\cite{as10} with this difference that here we formulate the latter using a 
more microscopic point of view.
It may be obtained from a GCM calculation in the following way,

\begin{equation}
\Psi^{cluster} = \int' dq f(q) \vert HFB(q) \rangle \vert_{N, Z,}
\end{equation}

\noindent
where the indices $N, Z$ indicate that the HFB wave function is projected 
on good particle number (neutrons and protons). The variable 
$q$ indicates the very asymmetric fission path of the $\alpha$ particle on its 
way to disintegration. In the end of the process, $q$ will correspond to the 
relative distance of the c.o.m. coordinates of the Pb core and 
the $\alpha$ particle. In fig.~\ref{pepin} we show a series of typical 
''Pb~+~$\alpha$'' shapes which $\vert HFB(q) \rangle$ should contain as a 
function of q (we agree that such shapes would be difficult to 
create technically: It is like a very 
asymmetric fission calculation). 

One should be aware of the fact that 
$\vert HFB(q) \rangle$ expressed in terms of excitations of the 
spherical $^{208}$Pb core + valence particles contains complicated many 
particle-many hole configurations (many quasi-particle configurations) 
such as, e.g. 5p-1h, 6p-2h,...9p-5h, etc., creating the cluster 
configuration. We put a prime on the integration sign to indicate that 
we took out of the sum 
the 4p-0h shell-model configurations, $\Psi^{SM}$, which we wrote formally 
apart in eq.~\ref{fo} in order to ease discussions below. 

After minimisation of energy with respect to the 
amplitudes $f(q)$ the corresponding GCM Schroedinger equation will supposedly 
contain a typical double-well potential for the $\alpha$ particle sitting 
on the right or left hand side of the Pb core.
 We suppose that the individual potential wells 
are deep enough to accomodate two states, the ground state and an excited 
state ($\nu$= 0 and 1). The excited state may be interpreted as a vibration of the $\alpha$ 
particle against the $^{208}$Pb core. 
The finite barrier will allow for tunnelling and the left-right 
degeneracy will be lifted to give rise to {\it four} different states with 
alternating parities, 
\begin{equation}\label{varphi}
\varphi^{\pm}_{0,1} = \frac{1}{\sqrt{2}}[\varphi^{R}_{0,1} \pm  
\varphi^{L}_{0,1}].
\end{equation}
The next step is to project the four wave functions on 
good spin value. We surmise that the small $\alpha$-grain on top of $^{208}$Pb 
perturbs sphericity only slightly. It will produce double-well scenarios 
weakly depending on spin. This naturally will produce for each spin  
quadruplets of states 
with alternating parities, i.e. four components $\Psi^{cluster}_{0,1}(I^{\pm})$ 
for eq.~\ref{fo}. Let us mention that spin projection also takes care of the $l=1$ 
mode where the $\alpha$ particle vibrates against the Pb core. 

These four cluster states still have to be mixed and orthogonalized with 
the SM configurations as indicated in eq.~\ref{fo}. Among the low-lying 
quadruplet of states belonging to the yrast band, only the positive-parity 
states will be affected. 
Actually the mixing may be relatively strong (it is predicted 
that the ground state contains about 30\% 
alpha-cluster configuration~\cite{va92}) pushing the other positive 
parity states up into energy regions where the highest member is yet 
to be detected. This also eventually explains why the natural order 
of +, -, +, - parities is changed to +, -, -, +. 
This picture qualitatively explains the experimental situation for 
the quadruplet of states of even spin comprising the ones of the yrast 
band (see fig.~\ref{States_212Po}a). 
One easily imagines that the projection on odd-spin values creates states at higher 
energy, since already at sphericity the odd-spin, negative-parity states 
are above the ones with even spins. Therefore, the whole scenario shown in 
fig.~\ref{States_212Po}
could find a natural explanation within the above lines of thought.

\begin{figure}[!h]
\includegraphics[width=4.5cm]{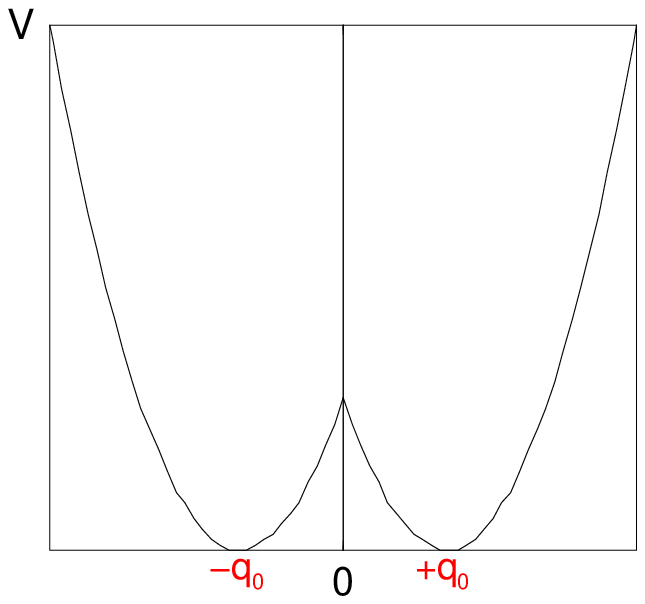}
\begin{center}
\includegraphics[width=8.5cm]{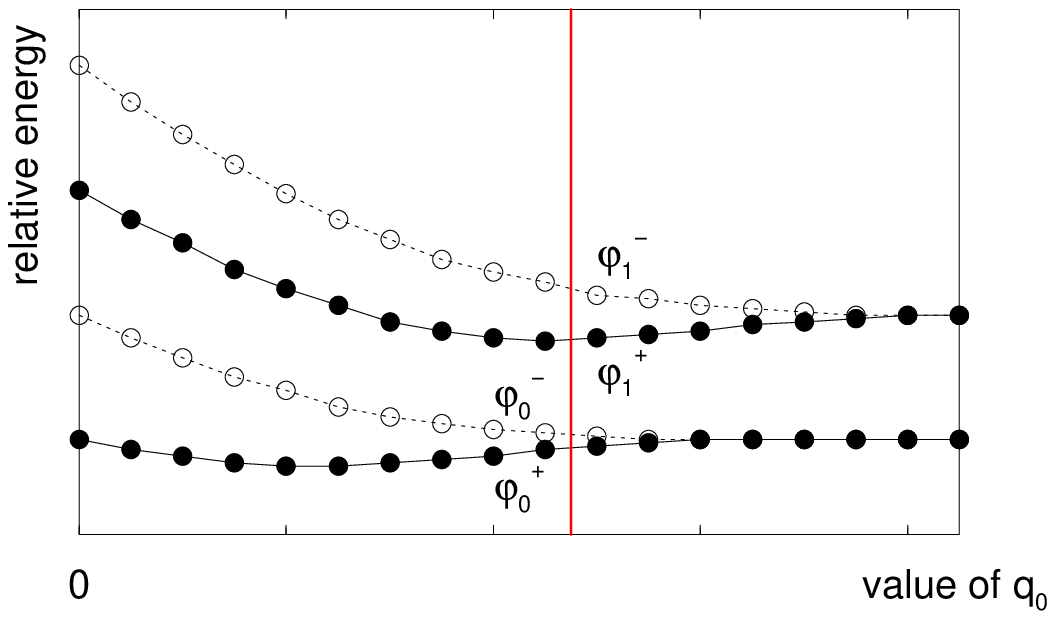}
\caption{{\bf Top}: $V(q)$ for a double-oscillator potential. 
{\bf Bottom}: Relative
energies, versus the distance between the two minima, for the four lowest
eigenstates of the double oscillator. When $q_0 =0$, the eigenvalues are those
of a simple harmonic oscillator. When the two wells are completely separated,
(large value of $q_0$), the energies of $\varphi^{\pm}_{0}$ are degenerated, as
well as those of $\varphi^{\pm}_{1}$. The vertical line indicates the four
energies obtained for a typical value of $q_0$.  
}
\label{pot_alpha}       
\end{center}
\end{figure}

To be slightly more specific, let us discuss a completely phenomenological
model imitating schematically the GCM calculation. The deformation 
Hamiltonian is then
\begin{equation}
H^{(I)}_{coll} = -\frac{\hbar^2}{2M} \frac{\partial^2}{\partial q^2} + V(q)
\end{equation}
\noindent
where $M$ is an adjustable inertia, depending eventually on spin and 
$V(q)$ is a double-well potential which for convenience may be taken as a 
double-oscillator potential, a typical textbook example~\cite{me61}, which is
sketched in the top part of fig.~\ref{pot_alpha}. The barrier penetration, which
is a function of the $q_0$ value, leads to the energy splittings of the  
$\varphi^{\pm}_{0}$ and $\varphi^{\pm}_{1}$ eigenstates (cf. eq.~\ref{varphi}): 
The relative energies 
of these four states are drawn in the bottom part of  fig.~\ref{pot_alpha}.  

The potential parameters also may depend on 
spin stemming from the spin projection. Using the dipole 
operator, as defined in eq.~\ref{dipolemoment}
and supposing that the cluster configuration of the even-parity yrast states
is admixed with 30 percent, we 
get a right order of magnitude for the $B(E1)$ values. Making the oscillator 
length of the potential slightly spin dependent, we also can reproduce the 
falling tendency of B(E1) vs $\Delta E(I)$ (see fig.~\ref{levels_BE1}) rather 
accurately.

The above scenario remains on a very schematic level. It should be 
verified by more microscopic studies which are planned for the future.
 
\subsubsection{Outlook}
The oscillatory motion of the $\alpha$-core distance around the 
equilibrium position is proposed for the first time. This would be a novel 
manifestation of $\alpha$-clustering, never noticed 
in light systems. In those cases, the $\alpha$-core system can rotate 
collectively about its center of mass (for a review of the so-called 
nuclear molecules, see~\cite{vo06}) and it is probably difficult to isolate 
an oscillatory motion, if any. 

In the chart of nuclides, there are two cases similar to $^{212}$Po: 
$^{104}_{~52}$Te as '$\alpha$+$^{100}$Sn' and $^{136}_{~52}$Te 
as '$\alpha$+$^{132}$Sn'. These $\alpha$-core systems fulfill  the two 
prerequisites:
\begin{itemize}
\item The ground state of the composite nucleus is above or close to 
the $\alpha$-decay threshold (cf. the {\it Ikeda} diagram displayed in 
fig.~1 of ref.~\cite{vo06}). 
\item The center of mass of the $\alpha$-core system is nearly merged 
with the center of the spherical core, so as the system cannot rotate 
collectively. 
\end{itemize}
Having the same value of dipole moment as $^{212}$Po, $^{136}$Te should 
exhibit enhanced electric dipole transitions.
On the other hand, the decay of the cluster states of $^{104}$Te should
be different since the E1 transitions are hindered as there is no shift 
between the center of charge and the center of charge, the two clusters 
having $N=Z$. Thus the $\alpha$ emission could compete for their decay.

One may speculate that adding more $\alpha$'s to the $^{208}$Pb core, 
like, e.g. two $\alpha$'s to give $^{216}$Rn, may exhibit similar 
physics. For example the two $\alpha$'s may move coherently as a 
$^8$Be and then the present scenario may repeat itself partially, 
or the two $\alpha$'s may move independently and, then, more complex 
structures can be expected.
\section{Summary and conclusions}\label{summary}

In summary, we have used the transfer of an $\alpha$-particle induced
by a heavy-ion beam at very low energy to populate excited states 
of $^{212}$Po. The level scheme has been extended up to $\sim 3.2$~MeV 
excitation energy. The $\gamma$ angular distributions
and $\gamma -\gamma$ angular correlations have been analyzed in order to 
assign spin and parity values to many observed states.
Several $\gamma$ lines with E$_\gamma <$ 1~MeV have been found to be 
strongly shifted and broadened by the Doppler effect, allowing for the 
measurements of the corresponding lifetimes by the DSAM method. The values, found 
in the range [0.1-0.6]~ps, lead to very enhanced E1 transitions. 

The excited states of $^{212}$Po can be distributed among two groups:
\begin{itemize}
\item The first one comprises all the states having natural parity. They bear
strong resemblance with those identified in the neighbouring isotopes and
isotones and can be explained in terms of single-particle excitations
dealing with the neutron and proton orbits lying close to the Fermi
levels, as well as their coupling to the 3$^-$ octupole vibration.
Nevertheless more stringent analysis reveals that 
several properties (such as B(E2) transition probabilities or 
$\alpha$-emission probabilities) are not accounted for by such an 
interpretation.
\item The second one comprises many states having non-natural parity which are
mainly grouped into 2 sets, 
a first one with even-$I$ values around 2~MeV excitation 
energy and a second one with odd-$I$ values
around 3~MeV. These levels only decay to the 
yrast states having the same $I$ value, by very enhanced E1 transitions 
(B(E1) $\sim 2\times10^{-2}$ -- $1\times10^{-3}$~W.u.). 
They are the fingerprints of the '$\alpha$+$^{208}$Pb' structure. 
\end{itemize}

Such an '$\alpha$+core' structure is observed for the first time. It could be 
also identified
in two other systems, $^{104}_{~52}$Te as '$\alpha$+$^{100}$Sn' 
and $^{136}_{~52}$Te as '$\alpha$+$^{132}$Sn'. 
While these two Te isotopes are located very far from 
the stability valley, they could be close at hand, thanks to the new 
worldwide facilities soon available.

{\bf Acknowledgement} 
The Euroball project was a collaboration between France, 
the United Kingdom, Germany,
Italy, Denmark and Sweden. A.A., P.P. and M.-G.P. are very 
indebted to their colleagues 
involved in the EB-02/17 experiment devoted to the fission fragments, 
in which the present data on $^{212}$Po were recorded. They thank 
the crews of the Vivitron, as well as M.-A.~Saettle
for preparing the Pb target, P.~Bednarczyk, J.~Devin, 
J.-M.~Gallone, P.~M\'edina and D.~Vintache
for their help during the experiment. 
We thank 
Dr. M. Mirea, Dr. W. Nazarewicz, Dr. W. von Oertzen, Dr. N. Rowley, 
and  Dr. T. Yamada for fruitful discussions.
This work was 
partially supported by the collaboration agreement between the
Bulgarian Academy
of Sciences and CNRS under contract No 16946, and by 
contract IDEI-119 of the Romanian Ministry of Education 
and Research.

\end{document}